\documentclass[journal,twoside,web]{IEEEtran}
\usepackage{generic}
\usepackage{cite}
\usepackage{amsmath,amssymb,amsfonts}
\usepackage{algorithmic}
\usepackage[pdftex]{graphicx}
\usepackage{textcomp}

\usepackage{colortbl}
\definecolor{grayTab}{gray}{0.95}

\usepackage{dirtytalk}

\usepackage{hyperref}
\hypersetup{
    colorlinks=true,
    linkcolor=blue,
    filecolor=magenta,      
    urlcolor=cyan,
}

\begin{document}

\title{Towards identifying optimal biased feedback for various user states and traits in motor imagery BCI}

\author{Jelena Mladenovi\'c, J\'er\'emy Frey, Smeety Pramij, J\'er\'emie Mattout, and Fabien Lotte
  \thanks{This work received support from Inria project lab BCI-LIFT and from the European Research Council with project BrainConquest (grant ERC\_2016\_STG\_714567).}
  \thanks{J. Mladenovi\'c was with Inria Bordeaux Sud-Ouest, LaBRI (CNRS, Univ. Bordeaux, INP), France and with Lyon Neuroscience Research Centre, COPHY team, (INSERM UMRS 1028, CNRS UMR 5292, Univ. Claude Bernard Lyon 1, F-69000) Lyon, France. She is now with School of Computing, Union University, Belgrade, Serbia (e-mail: jmladenovic@raf.rs).}
  \thanks{J. Frey is with Ullo, La Rochelle, France }
  \thanks{S. Pramij is with Inria Bordeaux Sud-Ouest, LaBRI (CNRS, Univ. Bordeaux, INP), France}
  \thanks{J. Mattout is with Lyon Neuroscience Research Centre, COPHY team, (INSERM UMRS 1028, CNRS UMR 5292, Univ. Claude Bernard Lyon 1, F-69000) Lyon, France}
  \thanks{F. Lotte is with Inria Bordeaux Sud-Ouest, LaBRI (CNRS, Univ. Bordeaux, INP), France }
  \thanks{Copyright (c) 2021 IEEE. Personal use of this material is permitted. However, permission to use this material for any other purposes must be obtained from the IEEE by sending an email to \mbox{pubs-permissions@ieee.org}.}
}

\maketitle

\begin{abstract}
\textit{Objective.} 
Neural self-regulation is necessary for achieving control over brain-computer interfaces (BCIs). This can be an arduous learning process especially for motor imagery BCI.
Various training methods were proposed to assist users in accomplishing BCI control and increase performance. Notably the use of biased feedback, i.e. non-realistic representation of performance.
Benefits of biased feedback on performance and learning vary between users (e.g. depending on their initial level of BCI control) and remain speculative.
To disentangle the speculations, we investigate what personality type, initial state and calibration performance (CP) could benefit from a biased feedback. \textit{Methods.} We conduct an experiment (n=30 for 2 sessions). The feedback provided to each group (n=10) is either positively, negatively or not biased. 
\textit{Results.} Statistical analyses suggest that interactions between bias and: 1) workload, 2) anxiety, and 3) self-control significantly affect online performance. For instance, low initial workload paired with negative bias is associated to higher peak performances (86\%) than without any bias (69\%). High anxiety relates negatively to  performance no matter the bias (60\%), while low anxiety matches best with negative bias (76\%).
For low CP, learning rate (LR) increases with negative bias only short term (LR=2\%) as during the second session it severely drops (LR=-1\%).
\textit{Conclusion.} We unveil many interactions between said human factors and bias. Additionally, we use prediction models to confirm and reveal even more interactions. 
\textit{Significance.} This paper is a first step towards identifying optimal biased feedback for a personality type, state, and CP in order to maximize BCI performance and learning.
\end{abstract}

\begin{IEEEkeywords}
Brain-computer interface (BCI), Electroencephalography (EEG), prediction models, feedback bias
\end{IEEEkeywords}

\section{Introduction}
\IEEEPARstart{A} Brain-computer interface (BCI) can be defined as \say{a system that measures neural activity and converts it into artificial output that replaces, restores, enhances, supplements, or improves natural neural output and thereby changes the ongoing interactions between the central nervous system and its external or internal environment} \cite{wolpaw2013brain}. 
Active BCIs require focus to produce mental commands towards an external device, and establish a new form of control (e.g. for movement \cite{milan2010invasive} or communication  \cite{farwell1988talking}).
To achieve moderate to high performances in active BCIs the user is supposed to understand the task and to produce distinct and stable brain signals \cite{lotte2018defining}. This is especially difficult for motor imagery MI-BCI, i.e., a BCI in which the machine is controlled by one's mental movements of limbs (e.g., hands, feet, tongue). To assist the user in accomplishing the task, BCI training approaches were proposed that apply various educational \cite{Keller2010}
and motivational theories \cite{Nakamura2002}. The training approaches can range from using social contexts, i.e., collaborative or competitive games \cite{Bonnet2013TwoImagery}, proprioceptive \cite{braun2016embodied}, immersive \cite{Alimardani2014EffectSystem.}, to gamified 3D environments \cite{Ron-angevin2009BraincomputerTechniques, mladenovic2017impact}, congruent with MI tasks \cite{christophe2018evaluation}, and many more \cite{roc2020review}. 
In addition, some investigated whether by influencing the belief about one's current performance can in fact affect their performance in reality, e.g., by providing biased (unrealistic) feedback  \cite{Barbero2010BiasedInterfaces., angulo2014effect, Alimardani2014EffectSystem., mladenovic2017impact}.

While some methods such as the use of 3D game versus 2D minimalist task managed to increase both performance and user experience significantly (on average) \cite{Ron-angevin2009BraincomputerTechniques}, other methods produced variable results because of their dependency on human factors \cite{mcfarland1998eeg, emami2020effects},  notably methods with biased feedback \cite{Barbero2010BiasedInterfaces., mladenovic2017impact}. Indeed, in \cite{Barbero2010BiasedInterfaces.} authors speculated that participants with initially high BCI control previously determined with calibration performance (CP) would not benefit from any type of biased feedback, while on the contrary low CP participants would. In \cite{mladenovic2017impact} the results were inconclusive about the impact of biased feedback on performance (not significant on average), while differently from \cite{gonzalez2011motor} where negative bias increased learning, in \cite{Alimardani2014EffectSystem., angulo2014effect} it did not. Namely, due to such variable results, authors in \cite{Alimardani2014EffectSystem.} suggested further investigation on the influence of personality types and user CP, while \cite{angulo2014effect} suggested personalizing the feedback bias.

On a side note, although the mentioned methods include biased feedback (and their results were compared in \cite{Alimardani2014EffectSystem.}), their approach and interface design vary and might be another reason for such variable results. For example, perceptually continuous alterations of the classification output in real-time which resulted in mild, adaptive feedback biases were provided in \cite{Barbero2010BiasedInterfaces.} and \cite{mladenovic2017impact}, while in \cite{gonzalez2011motor} and \cite{Alimardani2014EffectSystem.} authors performed absolute miss or match 30\% or 90\% of the time, as negative or positive feedback, respectively. Moreover, a 3D video game was presented in \cite{mladenovic2017impact}, a VR environment with proprioceptive feedback in \cite{Alimardani2014EffectSystem.}, a simple 2D (ball falling in basket) task in \cite{Barbero2010BiasedInterfaces.}, and discrete checkbox (correct/incorrect) feedback in \cite{gonzalez2011motor}. Nevertheless, we will not investigate these issues here, such differences in interface designs and their implications in performance variations are thoroughly discussed in \cite{jelena20standard}.

To propose training tasks or interfaces that fit each user, a variety of user traits, skills and states were identified as predictors to BCI performance. For instance, personality traits such as self-reliance positively correlates, while apprehension negatively correlates with performance \cite{JeunetPredictingPatterns}; and user contextual states such as motivation \cite{Kleih2010, hammer2012psychological}, attention \cite{grosse2011fronto}, confidence \cite{nijboer2008auditory}, sense of control \cite{Witte2013ControlTraining.},  and state of flow (a composite optimal state of control, immersion and pleasure) \cite{mladenovic2017impact} correlate positively to performance, while cognitive load correlates negatively with performance but only for those with performance below 75\% classification accuracy \cite{emami2020effects}. Additionally, competitiveness could be seen as a trait that predicts performance, as in sports there is evidence of a strong relationship between competitiveness, anxiety, self confidence and performance \cite{martin1991relationships}; competitive players showed to give more effort when compared to non-competitive players \cite{snyder2012virtual}, while the use of games and competitions to promote intrinsic motivation and performance has shown useful for learning various programming skills \cite{burguillo2010using}.

In this paper we wish to disentangle the speculations about the personality, states and CP involved in the success of biased feedback.  Using prediction models and standard statistical analyses, we investigate the joint effect of bias types (negative, positive and no bias)
and personality traits \cite{JeunetPredictingPatterns}, CP \cite{Barbero2010BiasedInterfaces.}, flow state \cite{mladenovic2017impact} and workload \cite{emami2020effects} on performance and learning. Prediction models can unveil which factors (traits, states, CP) individually predict performance and learning. Furthermore, they can reveal an interaction between those factors and the bias that in conjunction predict performance or learning. Positive significant interactions would prescribe the best pairs between bias and said factors that are associated with increased performance or learning. On the other hand, negative ones would suggest the worst pairs. 
Hence, this paper is an attempt at starting to build a guideline for matching bias to the user and maximize performance and learning.

In section \ref{sec:Experimental-DesignTux2}, we describe the experimental design, our bias function, signal processing and performance metrics, as well as our prediction models. In section \ref{sec:Preliminary-Results} we present our results including short comments; in \ref{sec:DiscussionTux2} we interpret more thoroughly our results, and finally in section \ref{sec:ConclusionTux2} we provide our concluding words, and describe future works and challenges for creating an adaptive model for MI-BCI based on these results. 

% HACK: avoid orphan heading
\begin{samepage}
\section{\label{sec:Experimental-DesignTux2}Materials and Methods}

\subsection{Experimental Design}
\end{samepage}

30 healthy participants were recruited (12 women, mean age: 28.56 years, SD: 6.96). The study was approved (validation number \emph{2019-05}) by the Inria ethics committee, the COERLE. We created a 3-condition-between-subject design: (1.) no\_bias
(control) group in which the classifier output is not biased, (2.) positive\_bias, and (3.) negative\_bias groups in which the classifier output is positively and negatively biased in real-time, respectively.

To ensure balance between groups according to performance, 10 participants were assigned to each group depending on the calibration scores in the first session.
We would determine if the current participant was on the low (50-65\%), medium (65-80\%) or high (80+\=) tier in terms of classification accuracy and assign them to a group depending on the constitution of all groups up to that point. Overall the average classification accuracy during the first session was 74.25\% in the negative\_bias group (SD=16.06), 66.38\% in the no\_bias group (SD=9.60) and 71.88\% in the positive\_bias group (SD=12.90). Note that 1-way ANOVA showed no significant difference between groups ($p=0.40$).

Each participant was engaged in 2 sessions (on 2 different days with a 1-to-5 days interval). For each 6-run session, participants played the Tux Racer game using left/right hand MI to move a virtual penguin Tux to the left/right respectively to catch fish. A run contained 40 trials (20 trials per MI class) and lasted 5 minutes and 25 seconds. Each 4 second long trial was visually marked by two flags on the racing course, within which a set of 8 closely arranged fish were to be caught within 3 seconds, see Figure \ref{fig:Screen-shot}. It was followed by a 4 second pause, during which controls were deactivated, and participants could rest or adjust their position if needed. Each session lasted around 2 hours and consisted of 3 parts.

\begin{figure}[h]
\begin{centering}
\includegraphics[width=0.9\columnwidth]{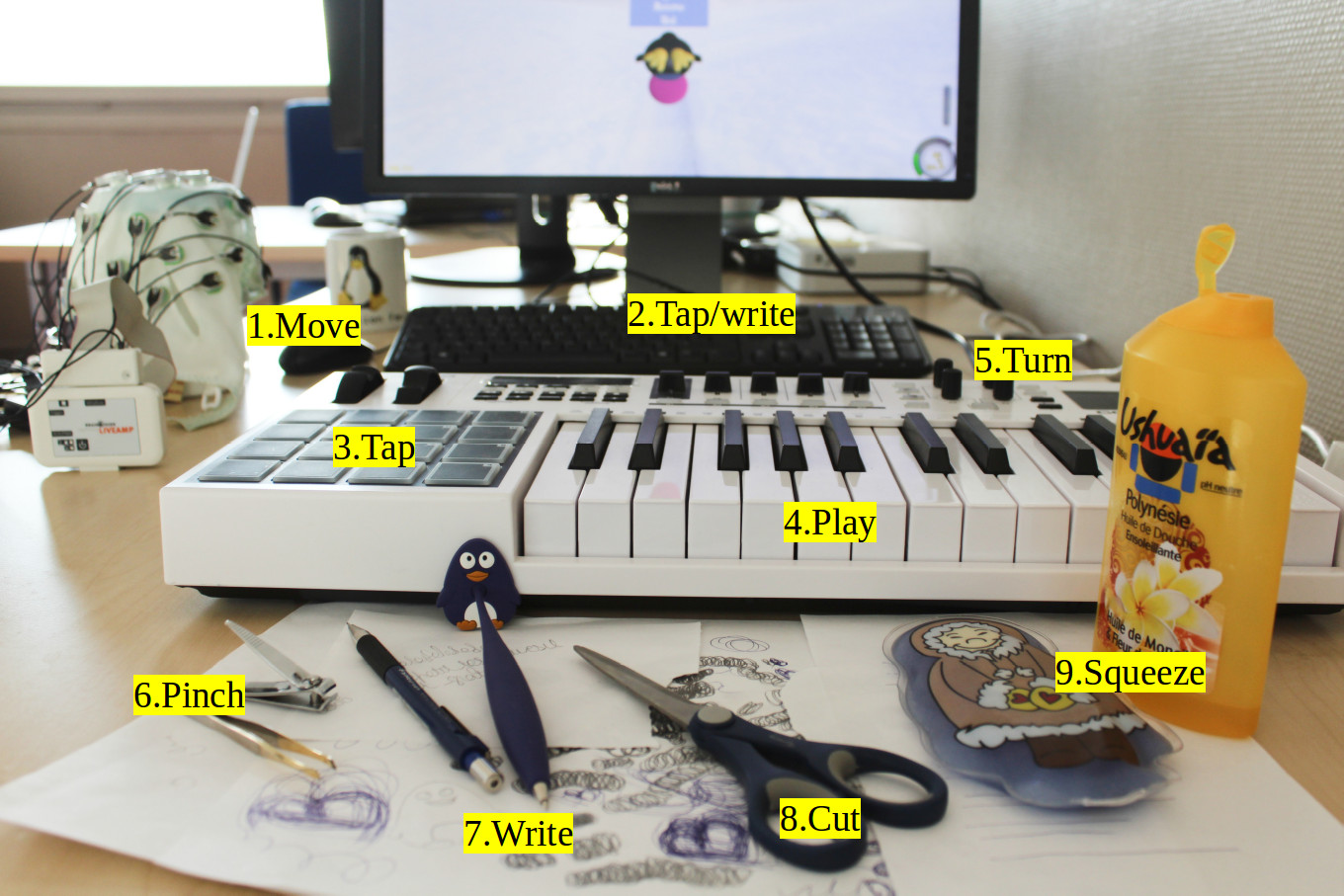}
\par\end{centering}
\caption{\label{fig:objects} Physical training on objects were proposed. By demonstrating the possible movements on such objects, novice users could have a better idea of how imagined movements can look or feel like.}
\end{figure}

Part I is the experiment preparation (around 30mins), during which participants:
(a) signed the consent form;
(b) filled in a demographics form;
(c) watched an explanatory 3-minutes animated movie about EEG, motor imagery, and the principles of the Tux Racer game;
(d) tried-out objects for motor training (Figure \ref{fig:objects}) during which the experimenter was placing the EEG cap. Participants were not obliged to try-out any of the presented objects unless they needed a somatosensory reminder of a movement, showed to increase performance of BCI novices \cite{Battison15_Effectiveness}. After the try-outs, they (e) observed their raw EEG signals with facial movements, jaw clenching and eye blinks in real-time to increase their sense of agency \cite{Vlek14_BCIAndUsersJudgmentOf}.

Part II is the calibration phase (2 runs of 40 trials each, around 15mins), during which participants:
(a) repeatedly performed their imagined movement for left or right hand at a time respectively with the left-right apparition of the fish;
(b) received ``sham'' feedback in which Tux was controlled by a script that generated quasi-random, rather optimistic behavior (participants were told that they were not the ones controlling the game);
(c) perceived Tux sliding on a pink snowboard, instead of its own belly, which was the visual indication to differentiate the calibration from the testing phase; and (d) finally, once the calibration runs were over, participants filled out a questionnaire on the subjective effort provided or workload (NASA-TLX) \cite{hart1988development} and flow state (EduFlow) \cite{Heutte2016}.

Part III is the testing phase (6 runs, around 45mins) during which the participants:
(a) controlled Tux with their left-right hand motor imagery;
(b) received either positively, negatively or not biased feedback;
(c) filled in NASA-TLX \cite{hart1988development} and EduFlow \cite{Heutte2016} questionnaires after each run;

Note that in each session, the machine was calibrated on training data from the same session only.

\paragraph{\textbf{Questionnaires}}

To assess the personality traits and affinities to competition, participants were asked to fill in the following questionnaires at home:

(1) Personality test 16PF5 \cite{cattell1995personality}, providing 16 primary scores of personality traits, and 5 global scores of personality (extroversion, anxiety, tough-mindedness, independence, and self-control), that are computed as linear combinations of the primary ones. Subsequent analysis used the 5 global scores.

(2) Revised Competitiveness Index \cite{houston2002revising}, from which we analyze only competition enjoyment scores.

To have a better understanding of one's experience and perceived difficulty during the experiment, the participants filled 2 questionnaires after each run:

(3) EduFlow questionnaire \cite{Heutte2016} provides 4 dimensions being
cognitive control, immersion, autotelism, and loss of self; the mean of all 4 dimensions we refer to as \emph{eduflow.} 

(4) NASA-TLX \cite{hart1988development} produces one score of workload (denoted as \emph{nasa\_score}).

\subsection{EEG Decoding and BCI Feedback}
\paragraph{\textbf{Signal Processing}}

The following equipment and software were used:
(i) A 32 channel Brain Products LiveAmp (wireless amplifier), (ii)
OpenViBE v2.2.0 \cite{Renard2010OpenViBE:Environments} for real-time signal processing in which we used: 
(1) Butterworth temporal filter, whose frequency band was optimized in a subject specific-way as in \cite{blankertz2007optimizing};
(2) Common Spatial Patterns (CSP) spatial filters to reduce the 32 original channels down to 6 ``virtual'' channels that maximize the differences between the two motor imagery classes \cite{Ramoser2000}; 
(3) Features computed as the EEG band power after CSP and band-pass filtering, using a 1s time window sliding every $1/16^{th}$ seconds;
(4) Probabilistic Support Vector Machine (SVM) with a linear kernel used to classify the features between left and right motor imagery. This way, the output of the SVM, $x_i\in[0,1]$, indicates both the class estimate and the confidence associated with that estimation. 

CSP filters were trained prior to the classifier over the two runs of the calibration phase of each session. As per \cite{Ramoser2000}, the process automatically selected weights so that the variances are optimal for the discrimination of the left/right classes (variance is maximized between classes and minimized within classes). The 3 pairs of filters selected after this process corresponded to the 3 largest and 3 lowest eigenvalues. The same frequency band was fed to the CSP and then to the SVM. We did not perform hyperparameters fitting for the SVM, leaving to default the parameters exposed by the LIBSVM\footnote{\url{https://www.csie.ntu.edu.tw/~cjlin/libsvm/}} implementation included in OpenViBE (C-SVC type, ``cost'' set to 1.0, shrinking enabled).

\paragraph{\textbf{Game controls}}

We used the Lab Streaming Layer (LSL\footnote{\url{https://github.com/sccn/labstreaminglayer}}) to control a virtual joystick that in turn controls the penguin. This means that the classifier output from OpenViBE was streamed via LSL in real-time and linearly mapped onto the -45 to 45 degrees angle range the penguin would take. In between trials the controls were deactivated so that Tux could not be manipulated and participants could rest.

\paragraph{\textbf{Game design}}

\begin{figure}[h]
\begin{centering}
\includegraphics[width=0.7\columnwidth]{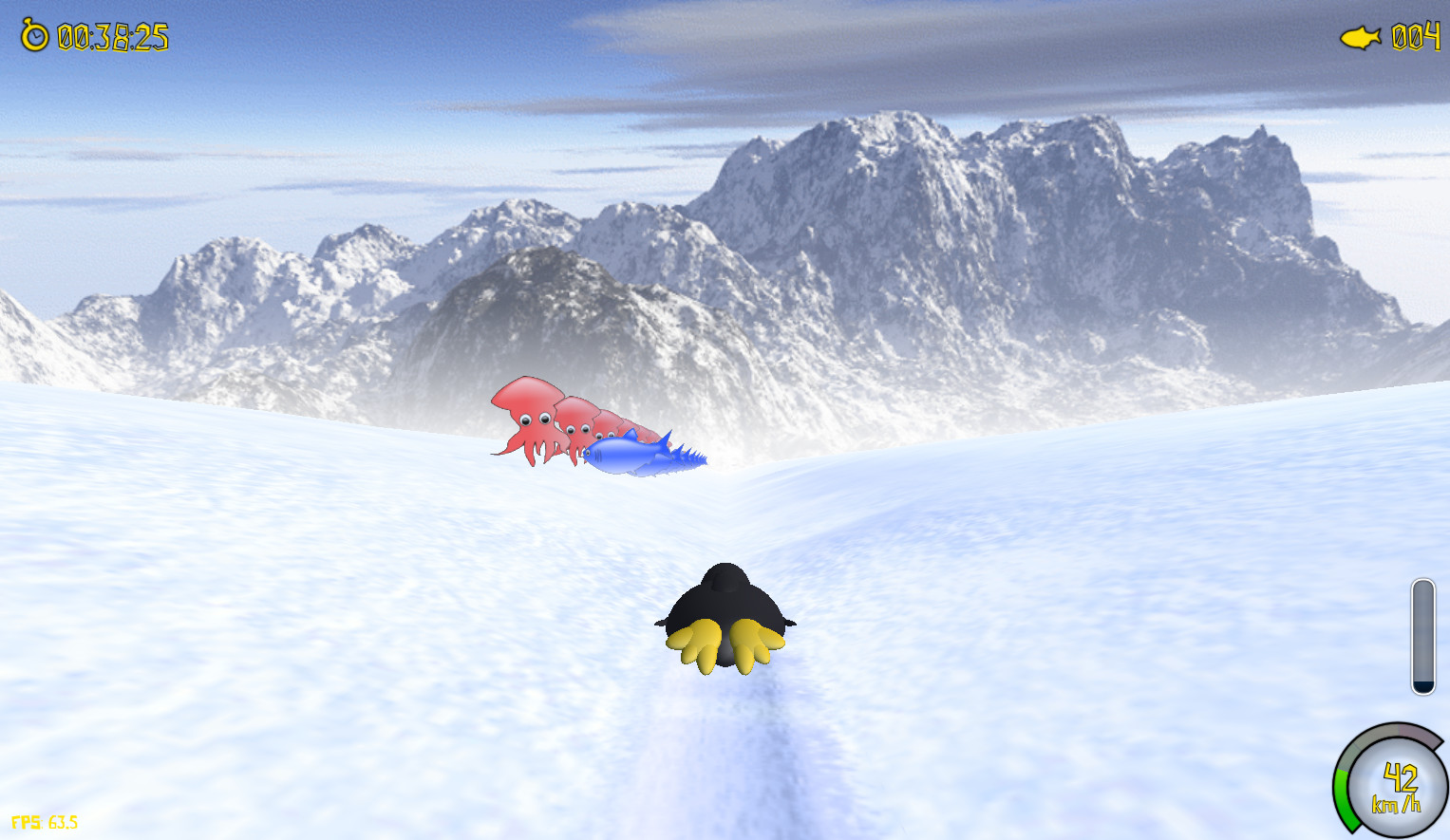}
\par\end{centering}
\caption{\label{fig:Screen-shot}BCI Tux Racer game during a trial for left hand MI.
}
\end{figure}

We modified\footnote{\url{https://github.com/jelenaLis/tux-modifs}} the open-source Tux Racer game by creating a bobsleigh ski-course through which Tux would slide in, at constant speed. We fixed the speed so that each run would last the same amount of time. Tux would then naturally slide back to the center of the course between each trial as the controls were deactivated. We placed one row of fish close to the center (each fish providing 1 point), and a row of squids close to extremities (each squid providing 5 points). The collected points are displayed in the upper right corner, the speed and indication of position in the course are in the bottom right corner, while the time elapsed is displayed in the upper left corner, see Figure \ref{fig:Screen-shot}.

\paragraph{\label{Mean-re-centering}\textbf{Mean re-centering}}

To account for the data non-stationarity, (i.e., EEG data statistics varying over time which may put the classifier out of center), we implemented the method proposed in \cite{vidaurre2010toward}. It consists in recentering the classifier (here SVM) hyperplane over time to avoid any inclination towards one class. This way we can assure the validity of our positive/negative bias strategies.
We applied this method by calculating the mean \emph{c} of
the classifier output for each run. As left-right classes are balanced the mean should be zero. If $c \neq 0$
we then subtract that amount from the classifier output to center it around zero, i.e., this value becomes the new mean. 

We perform re-centering after each run during testing, and calculate the mean from only the previous run (without including the calibration run). As there are many issues that could happen during a run, for instance, a faulty electrode would impact the classifier to choose one class throughout the whole run and would cause the hyperplane to undergo a drastic movement for the wrong reason, we bounded the mean estimation $c\in\left[-0.5,0.5\right]$, i.e., if $c>0.5$ then \emph{c = 0.5} or $c < -0.5$ then \emph{c = -0.5}. This method is implemented in every group, including the no\_bias group. 

\paragraph{\label{par:Bias-function.}\textbf{Bias function}}

\begin{figure}[h]
\begin{centering}
\includegraphics[width=0.77\columnwidth]{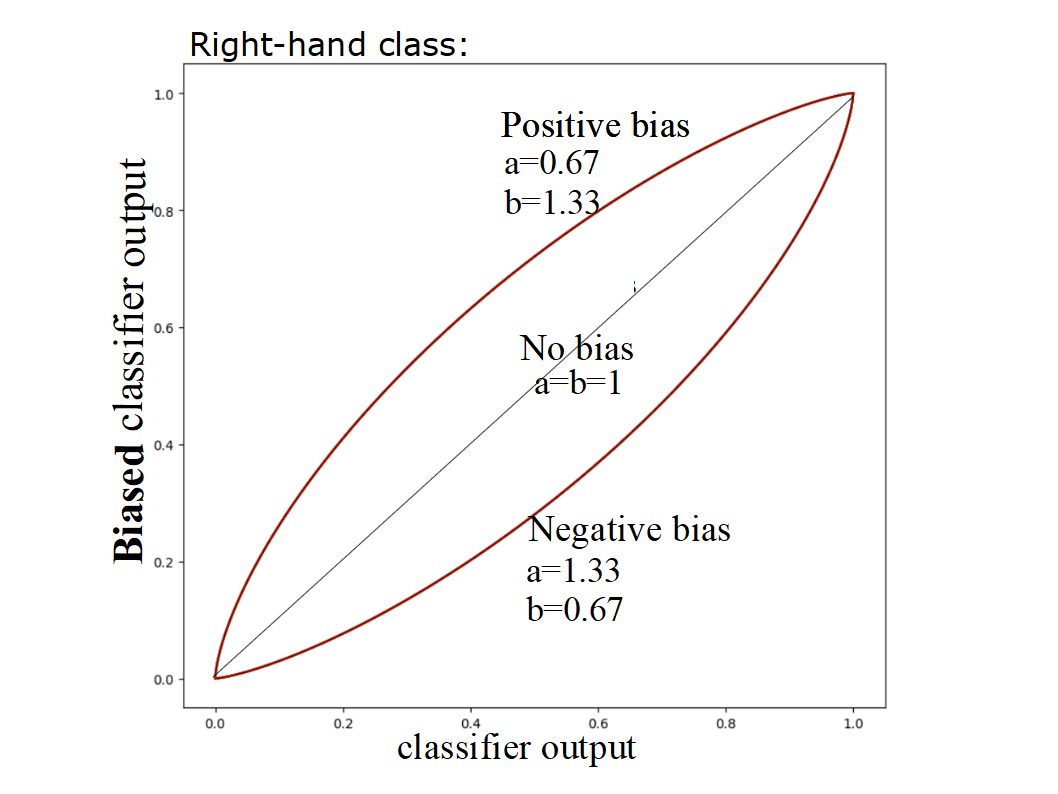}
\par\end{centering}
\caption{\label{fig:rainbow}Example for the right-hand class bias CD function: negative, positive and no\_bias, in which we map the classifier output (\emph{x-axis}) to the biased output (\emph{y-axis}).}
\end{figure}

We used the beta cumulative distribution function (beta CDF) to map the classifier output (which are probabilities) to a biased classifier output (every 16$^{th}$ of a second). This function contains 2 parameters \emph{a} and \emph{b} which enable us to easily control the slope, and create any kind of bias. Notably, from the identity function $a=b=1$ (used for the no\_bias group), we add and subtract a shift value \textit{k} for our bias function. The parameters for positive bias are $a_P=1 - k$, and $b_P=1 + k$, while for negative bias are $a_N=1 + k$ and $b_N=1 - k$, see Figure \ref{fig:rainbow}. K-value is chosen empirically \emph{k = 0.33} based on the flow theory (most recent one in motivational psychology \cite{Nakamura2002}), and serves to avoid boredom as well as frustration.
Note that positive and negative bias functions are mutually symmetric, but individually are not. 

From our previous study \cite{mladenovic2017impact, mladenovic2019computational} we noticed the drawbacks of a simple linear function as bias. Indeed, it could happen that participants verify their control over the system by performing the opposite hand motor imagery. That is, they would push Tux away from the fish contrary to the task at hand, just to verify if it is not sham feedback. With the simple linear bias function Tux was limited to only one side of the ski-course, and consequently the participants speculated a biased feedback. This phenomenon reoccurred in this study as well. Few participants admitted through informal interviews their need to explore and ``break the rules''. Fortunately, with the new bias function, users were able to attain minimal classification output as well as maximal one at all times, and they did not have any speculation about the feedback.

\subsection{Evaluation Metrics}

\paragraph{\textbf{Performance}}
Online performance corresponds to the peak and average performance of the classifier that controlled the video game, i.e. the highest and the mean classification accuracy over all trials' time windows, respectively. 
CP was computed offline with a 5-folds cross validation of the concatenated 2 runs recorded during calibration phase. Note that online performance comprehend ``real'' BCI performance of the user, before applying any bias.

\paragraph{\textbf{Learning}}
Learning rate (LR) is the slope (coefficient) of the linear regression of online performance over runs within a session, e.g., below zero indicates a decrease in learning while above indicates an increase. LR is here session sensitive (1 value per session), as a person may not linearly continue learning from the end of the first session to the beginning of the second one. Thus, this metric is useful for observing changes between sessions, however it is less evident to predict. For this reason, in our prediction model we use a simple difference in performance between sessions, denoted as \textit{progress}.

\paragraph{\textbf{Baseline}}
During the calibration phase of each session we assessed 1 value of flow, workload and CP. As we re-calibrate the system for each session, for subsequent analyses we used the assessed values (1 per session), denoted with suffix "\_baseline". However, to avoid learning effect for progress prediction, we use only the first session values, denoted with suffix "\_reference" (1 per participant). Thus baseline values are per session, while reference ones are per participant.

\paragraph{\textbf{High/low factors}}
To have a better understanding of the interaction between bias and human factors, we  separate  the factors into high  and  low  using  the median  value  within  a  particular trait or state, i.e. high/low (anxiety, extroversion, independence, tough-mindedness, self-control), and high/low (workload, flow, CP).

\subsection{\label{par:elasticNet}Prediction models}

Let online performance be $Y^{n}$ for \emph{n} subjects, using $X^{n\times p}$ covariates for \emph{p} predictive factors. We are interested in the interactions between three bias types $r_k \in \mathbb{B}, \, k=0,1,2$ on one hand, and traits (5 global personality traits, competition enjoyment), baseline states (CP, flow and workload "\_baselines"), $s_j \in \mathbb{R}, \, j=1,..9 $ on the other.
Finally, the interaction between $r_k$ and $s_j$ or product  $r_k s_j = x_{9k+j}$ yields $p=27$ predictors $X^{n\times p}$.
In order to find predictive models, we used  Elastic-Net regression \cite{zou2005regularization}, which combines both Ridge (Tikhonov regularization) and Lasso (Least Absolute Shrinkage and Selection Operator). This method is often used in other fields when $p >> n$ \cite{engebretsen2019statistical}.
We implemented Elastic-Net regression as follows.

\[
\beta_{net}=\underset{\beta\in \mathbb{R}^{p}}{argmin}(\Vert Y-X\beta\Vert_{2}^{2}+\lambda[\frac{1}{2}(1\text{-}\alpha)||\beta||_{2}^{2}+\alpha||\beta||_{1}])
\]

Where $\beta$ represent regression weights, $\lambda$ is the regularization or shrinkage parameter, while the parameter \emph{$\alpha$ } balances the respective contribution of the two penalty terms $\Vert\beta\Vert_{2}^{2}$ and  $\Vert\beta\Vert_{1}$. Strictly speaking, $\Vert.\Vert_{2}^{2}$ indicates the $l_{2}$ norm 
while $\Vert.\Vert_{1}$ indicates the $l_{1}$ norm. 
If \emph{$\alpha=0$} then it boils down to Ridge regression with $l_{2}$
penalty, otherwise if \emph{$\alpha=1$} it amounts to Lasso regression with $l_{1}$ penalty. Note that the offset term is incorporated into X.

Two or more predictors are said to interact if their combined effect is different (lesser or greater) than the sum of the effect of each predictor taken separately. 
Then our model for predicting performance $y_i$ for each subject \textit{i}:

$$ y_i = \beta_0 + \sum_{k=0}^2 \,\, r_{i,k}\,\, \sum_{j=1}^{9} \,\, \beta_{9k+j} \,\, s_{i,j}$$

where  $\beta_{9k+j}$ are the coefficients (from 1 to 27) of the interactions $r_{i,k}$ and $s_{i,j}$. Knowing that $r_{i,k} \in \mathbb{B}$ is a Boolean, categorical predictor,
there exists only one \textit{k} among 0, 1 and 2 such that $r_{i,k} = 1$.
In other words, there can be only one bias type per participant $i$, the other two are equal to zero.
The same procedure is performed for the prediction of progress, except that $s_{i,j}$ baseline states or "\_baselines" are replaced with "\_references", e.g. $s_{i,1}$ is CP reference.

To optimize the prediction model, we perform a Leave One Subject Out (LOSO) cross-validation using the R function cva.glmnet from glmnetUtils\footnote{\url{https://cran.r-project.org/package=glmnetUtils}}, which uses a coordinate descent algorithm that, for each $\alpha\epsilon[0,1]$ (10 values by default), generates an entire path of solutions in $\lambda$ (100 values).
To avoid over-fitting, we performed a nested cross-validation LOSO to have different sets for training and testing during parameter search.
We hence performed a n-fold cross validation in the outer loop for model evaluation and a (n-1)-fold cross validation in the inner one for hyperparameter ($\alpha, \lambda$) selection.

In order to asses the prediction power of our models we compared their performance against random (null) models, where we kept hyperparameter values and users' factors but shuffled the outcome variable (either performance or progress). We did so using two methods. In the first method, similar to \cite{candia2019enetxplorer}, we shuffled outcome variables for both train and test sets. In the second method we shuffled data for the training set only, and tested the random models on real outcome variables. For both methods LOSO cross-validation was performed, shuffling the data a thousand times per fold. We could thus compare statistically the observed error rate across participants with error rates derived from our null hypothesis. We performed two tests because there is no consensus in the literature as about what is the best way to build random models. However, since both methods yielded the same results, we report only one p-value in the rest of the paper.

To get the final coefficients of the regression, we ran the model using the selected parameters \emph{$\alpha$} and \emph{$\lambda$} on the whole dataset. To estimate the variance, we computed the standard deviation of the selected coefficients during a LOSO cross-validation. Finally, to assess the significance of model's features, it was compared to a random model made of a thousand permutations using R package eNetXplorer \cite{candia2019enetxplorer}. This test based on permutation is robust to repeated measurements, in our case the several performance scores per subject.

For practical reasons, we predict online performance per participant using data from each run (12 values per participant).  We indeed used performance over runs so that eNetXplorer can compute enough permutations to test the significance with $p<0.01$.
Nevertheless, almost identical coefficients and RMSE were found for the model of performance averaged per session. Results' similarities confirm the insensitivity of permutations to dependent (within subject) factors. However, for the sake of readability, we do not report those model's results.

\subsection{Statistical Analysis}
Our statistical analysis involves ANOVA tests for determining the differences between our 3 bias groups. All the corresponding p-values were corrected for multiple comparisons with false discovery rate (FDR) \cite{Noble2009}, with the threshold being set to $p<0.05$. To avoid repetition, all results are presented after the p-values were corrected with FDR.
As for evaluating the validity of prediction models, we use the standard metric such as root mean squared error (RMSE), and the coefficient of determination $R^2$ (adjusted for multiple predictors).

\section{\label{sec:Preliminary-Results}Results}

\subsection{Group balance verification}

When performing 1-way ANOVA (independent variable: group, dependent: either calibration reference, dimensions of EduFlow score from calibration, NASA-TLX score from calibration, or personality traits) there is no significant difference between groups.
This could indicate that the groups were balanced considering their baseline performance, flow states and workload at the beginning of the experiment.

\subsection{Differences between groups within sessions}

In the following, we use workload, eduflow and CP from both calibration phases (nasa\_baseline, flow\_baseline, calibration\_baseline). 

\paragraph{\textbf{Performance}}

When performing 3-way ANOVAs (independent variables: group, session and high/low factors, dependent variable: either peak or average performance),
there is a significant effect ($p<0.001$) of high/low calibration performance on both peak and average online performance, meaning that e.g. low calibration performance would yield low online performance, no matter the bias. 

\begin{figure}[h]
\begin{centering}
 \includegraphics[width=\columnwidth]{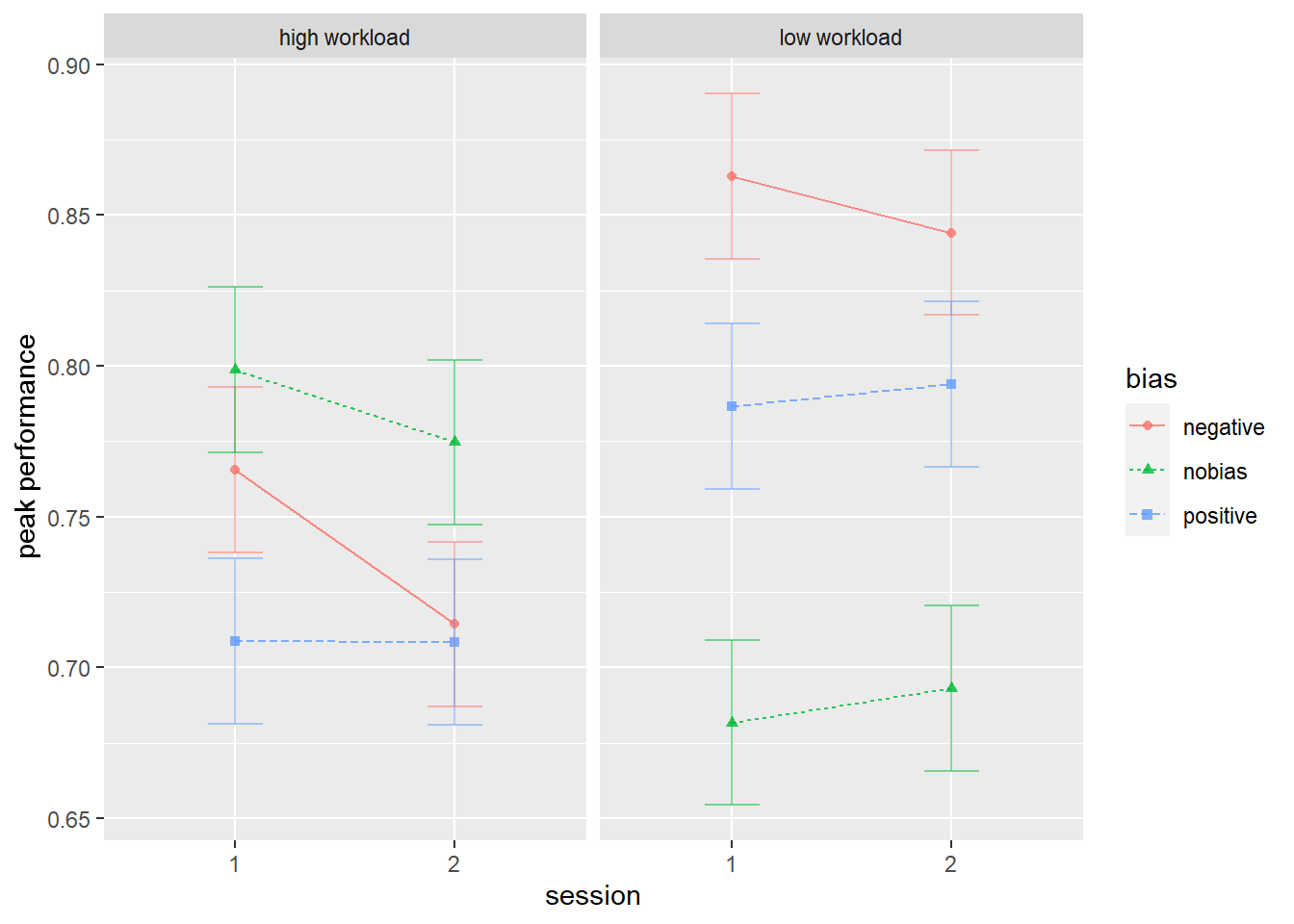}
\par\end{centering}
\caption{\label{PeakNasa} Significant interaction between high/low workload and bias (p$<$0.05) for peak performance within 2 sessions.
}
\end{figure}

We found a significant interaction: bias$\times$high/low workload, for both peak ($p=0.036$), and average performance ($p=0.022$), see Figure \ref{PeakNasa} and Table \ref{tab:stats}. This suggests that participants with initially low workload (low NASA\_reference) benefit most from a negative bias (peak $\simeq85\%$) and least from no\_bias (peak $\simeq69\%$). On the other hand, high workload participants do not require any bias to gain reasonably high performances (peak $\simeq79\%$).

\begin{figure}[h]
\begin{centering}
\includegraphics[width=\columnwidth]{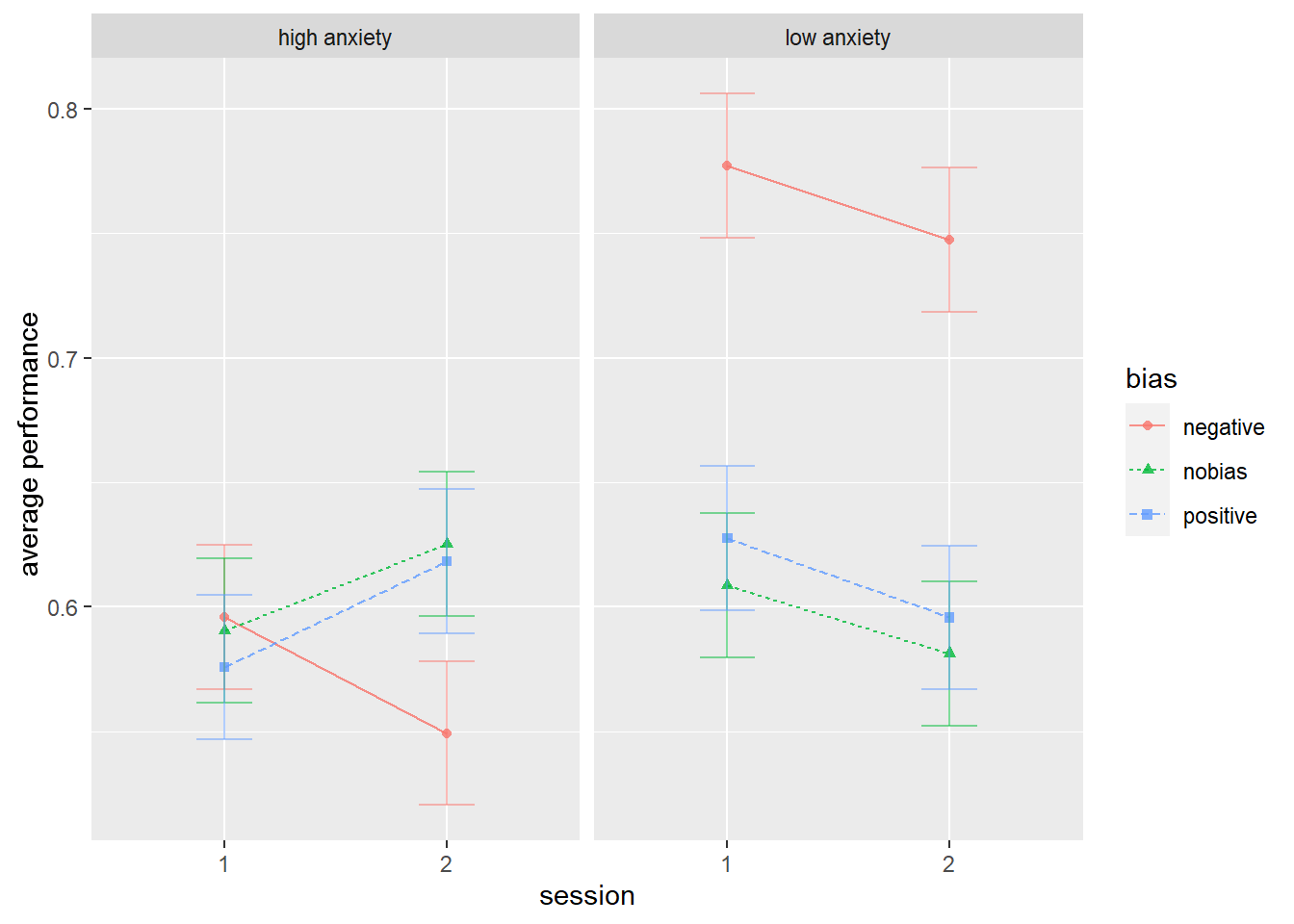}
\par\end{centering}
\caption{\label{AvgAnxiety} Significant interaction between high/low anxiety and bias (p$<$0.05) for average performance within 2 sessions.}
\end{figure}

There was a significant interaction: bias$\times$high/low anxiety for average ($p=0.031$, see Figure \ref{AvgAnxiety}), and a tendency for peak performance ($p=0.080$). This would suggest that low anxiety participants would benefit most from a negative bias (avg $\simeq76\%$) and least from no\_bias (avg $\simeq60\%$). On the other hand, highly anxious participants have low performance no matter the bias (avg $\simeq59\%$).

\begin{figure}[h]
\begin{centering}
\includegraphics[width=\columnwidth]{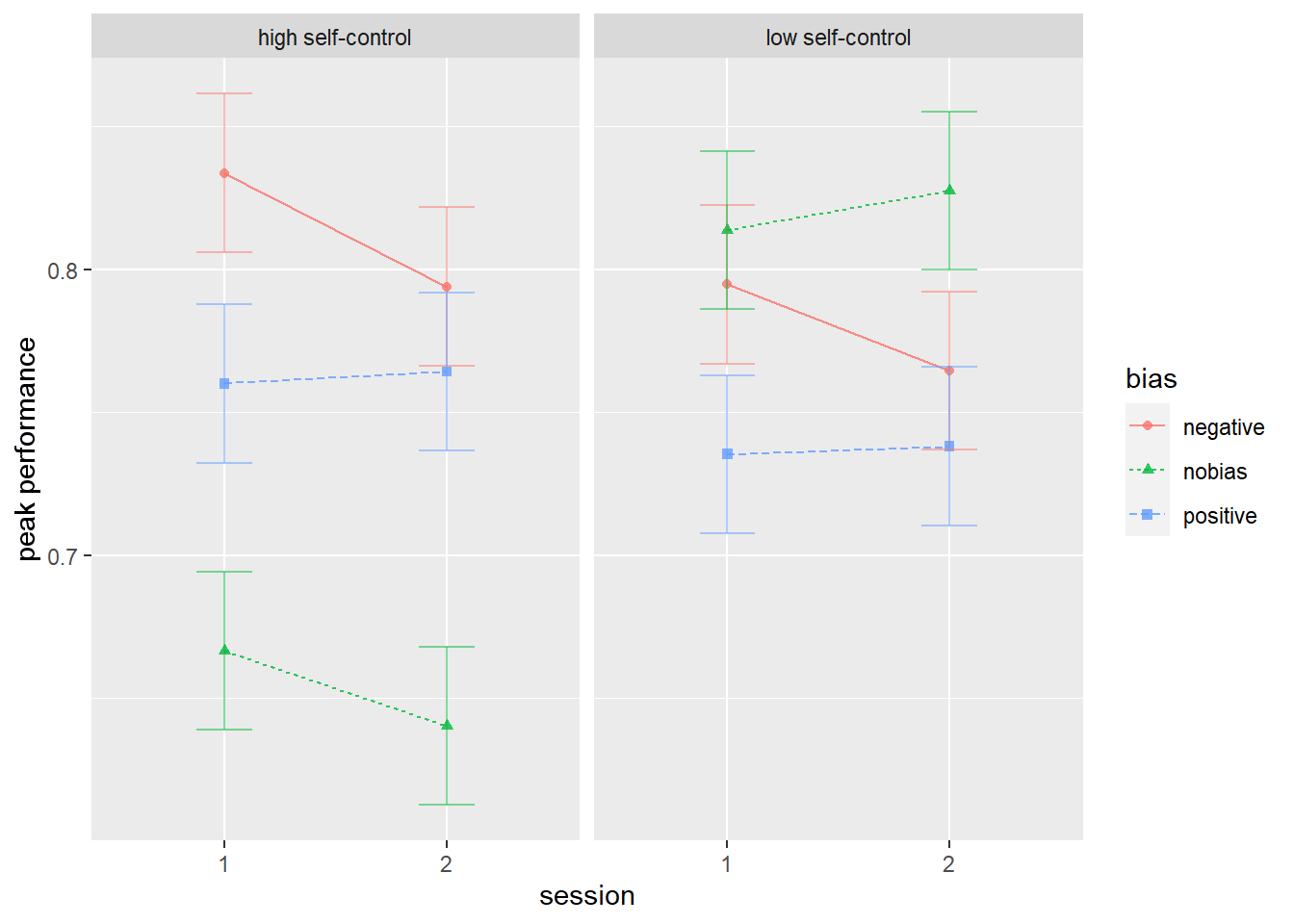}
\par\end{centering}
\caption{\label{PeakSelfcontrol} Significant interaction between high/low self-control and bias (p$<$0.05) for peak performance within 2 sessions.}
\end{figure}

Finally, there was a significant interaction: bias$\times$high/low self-control for average ($p=0.037$), and for peak performance ($p=0.042$), see Figure \ref{PeakSelfcontrol}. This result suggests that high self-controlled participants have higher performances with negative bias (peak $\simeq81\%$), than with no\_bias (peak $\simeq65\%$) which is the inverse for low self-controlled participants.

% the big table
\paragraph{\textbf{Learning rate}}

\begin{figure}[h]
\begin{centering}
\includegraphics[width=\columnwidth]{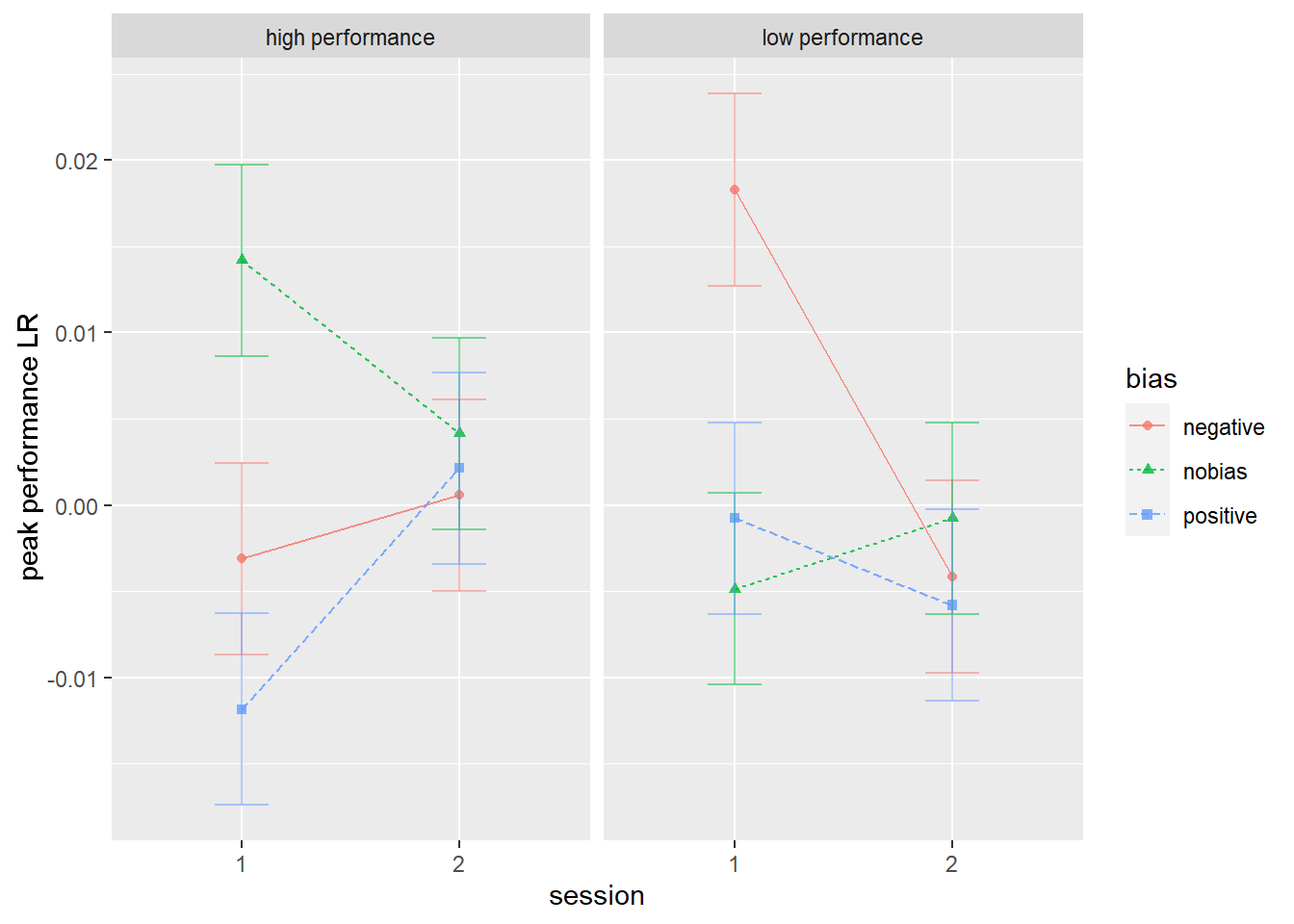}
\par\end{centering}
\caption{\label{fig:learnRatePerformers} Significant interaction between high/low calibration performance and session ($p<0.05$) and bias all together ($p<0.01$) for learning rate of peak performance.}
\end{figure}

With a 3-way ANOVA (independent variable: group (bias), session, high/low factor; dependent variable: learning rate) we get significant interactions: bias$\times$session (p=0.05), high/low CP$\times$session ($p<0.05$); and high/low CP$\times$bias$\times$session ($p<0.01$), see Figure \ref{fig:learnRatePerformers} and Table \ref{tab:stats}. This result indicates that CP, and bias can directly influence learning within sessions.

Negative bias increases learning rate within first session for those with low CP, but already in the second session it severely impedes learning rate, from LR $\simeq1.8\%$ to LR $\simeq-0.4\%$.
Positive bias seems to have a generally negative influence on learning (mostly below zero).
While no\_bias seems to be the best fit for high CP, although for only one session (from LR~$\simeq~1.4\%$ to LR~$\simeq~0.4\%$).

To have an overview of the results, we summarize all the peak and average performances, and learning rates between groups and sessions, see Table \ref{tab:stats}.

\begin{table*}[h]
\tiny
\centering
\begin{tabular}{lllllll}
\multicolumn{7}{c}{\textbf{Peak Performance (SD)}}                                                                                                                                                                                                                                                                                                                                                                                        \\
\rowcolor{grayTab} 
Average                                                     & \multicolumn{6}{l}{\cellcolor{grayTab}76.1 (11.7)}                                                                                                                                                                                                                                                                                                                     \\
\multicolumn{1}{l|}{}                                       & \multicolumn{2}{c|}{\textbf{Negative bias}}                                                                             & \multicolumn{2}{c|}{\textbf{Positive bias}}                                                                               & \multicolumn{2}{c|}{\textbf{No bias}}                                                                                  \\
\rowcolor{grayTab} 
\multicolumn{1}{l|}{\cellcolor{grayTab}Average}        & \multicolumn{2}{l|}{\cellcolor{grayTab}79.5 (11.5)}                                                                & \multicolumn{2}{l|}{\cellcolor{grayTab}74.9 (10.2)}                                                                  & \multicolumn{2}{l|}{\cellcolor{grayTab}73.7 (12.5)}                                                               \\
\multicolumn{1}{l|}{}                                       & \multicolumn{1}{c|}{\textbf{Session 1}}                   & \multicolumn{1}{c|}{\textbf{Session 2}}                     & \multicolumn{1}{c|}{\textbf{Session 1}}                     & \multicolumn{1}{c|}{\textbf{Session 2}}                    & \multicolumn{1}{c|}{\textbf{Session 1}}                   & \multicolumn{1}{c|}{\textbf{Session 2}}                    \\
\rowcolor{grayTab} 
\multicolumn{1}{l|}{\cellcolor{grayTab}Average}        & \multicolumn{1}{l|}{\cellcolor{grayTab}81.2 (11.8)}  & \multicolumn{1}{l|}{\cellcolor{grayTab}77.9 (11.1)}    & \multicolumn{1}{l|}{\cellcolor{grayTab}74.8 (9.90)}    & \multicolumn{1}{l|}{\cellcolor{grayTab}75.1 (10.7)}   & \multicolumn{1}{l|}{\cellcolor{grayTab}74.0 (11.8)}  & \multicolumn{1}{l|}{\cellcolor{grayTab}73.4 (13.2)}   \\
\multicolumn{1}{l|}{Calibration High}                          & \multicolumn{1}{l|}{89.2 (7.62)}                          & \multicolumn{1}{l|}{86.8 (6.91)}                            & \multicolumn{1}{l|}{81.9 (7.88)}                            & \multicolumn{1}{l|}{80.6 (10.1)}                           & \multicolumn{1}{l|}{76.7 (13.5)}                          & \multicolumn{1}{l|}{77.4 (13.9)}                           \\
\multicolumn{1}{l|}{Calibration Low}                          & \multicolumn{1}{l|}{73.4 (9.68)}                          & \multicolumn{1}{l|}{69.0 (6.32)}                            & \multicolumn{1}{l|}{67.6 (5.55)}                            & \multicolumn{1}{l|}{69.6 (8.16)}                           & \multicolumn{1}{l|}{71.3 (9.30)}                          & \multicolumn{1}{l|}{69.4 (11.4)}                           \\
\rowcolor{grayTab} 
\multicolumn{1}{l|}{\cellcolor{grayTab}Workload High}     & \multicolumn{1}{l|}{\cellcolor{grayTab}76.6 (9.80)}  & \multicolumn{1}{l|}{\cellcolor{grayTab}71.4 (9.71)}    & \multicolumn{1}{l|}{\cellcolor{grayTab}70.9 (8.91)}    & \multicolumn{1}{l|}{\cellcolor{grayTab}70.8 (8.69)}   & \multicolumn{1}{l|}{\cellcolor{grayTab}79.9 (11.8)}  & \multicolumn{1}{l|}{\cellcolor{grayTab}77.5 (13.4)}   \\
\rowcolor{grayTab} 
\multicolumn{1}{l|}{\cellcolor{grayTab}Workload Low}     & \multicolumn{1}{l|}{\cellcolor{grayTab}86.0 (11.9)}  & \multicolumn{1}{l|}{\cellcolor{grayTab}84.4 (8.40)}    & \multicolumn{1}{l|}{\cellcolor{grayTab}78.7 (9.43)}    & \multicolumn{1}{l|}{\cellcolor{grayTab}79.4 (10.9)}   & \multicolumn{1}{l|}{\cellcolor{grayTab}68.2 (8.66)}  & \multicolumn{1}{l|}{\cellcolor{grayTab}69.3 (11.9)}   \\
\multicolumn{1}{l|}{Anxiety High}                              & \multicolumn{1}{l|}{73.4 (9.68)}                          & \multicolumn{1}{l|}{69.0 (6.32)}                            & \multicolumn{1}{l|}{72.7 (10.3)}                            & \multicolumn{1}{l|}{73.7 (12.0)}                           & \multicolumn{1}{l|}{74.2 (9.79)}                          & \multicolumn{1}{l|}{75.4 (13.1)}                           \\
\multicolumn{1}{l|}{Anxiety Low}                              & \multicolumn{1}{l|}{89.2 (7.62)}                          & \multicolumn{1}{l|}{86.8 (6.91)}                            & \multicolumn{1}{l|}{76.8 (9.26)}                            & \multicolumn{1}{l|}{76.6 (9.07)}                           & \multicolumn{1}{l|}{73.9 (13.7)}                          & \multicolumn{1}{l|}{71.4 (13.3)}                           \\
\rowcolor{grayTab} 
\multicolumn{1}{l|}{\cellcolor{grayTab}Self-control High} & \multicolumn{1}{l|}{\cellcolor{grayTab}82.9 (13.6)}  & \multicolumn{1}{l|}{\cellcolor{grayTab}79.4 (11.9)}    & \multicolumn{1}{l|}{\cellcolor{grayTab}76.0 (11.7)}    & \multicolumn{1}{l|}{\cellcolor{grayTab}76.4 (11.0)}   & \multicolumn{1}{l|}{\cellcolor{grayTab}66.7 (6.44)}  & \multicolumn{1}{l|}{\cellcolor{grayTab}64.0 (4.64)}   \\
\rowcolor{grayTab} 
\multicolumn{1}{l|}{\cellcolor{grayTab}Self-control Low} & \multicolumn{1}{l|}{\cellcolor{grayTab}79.5 (9.67)}  & \multicolumn{1}{l|}{\cellcolor{grayTab}76.5 (10.3)}    & \multicolumn{1}{l|}{\cellcolor{grayTab}73.5 (7.69)}    & \multicolumn{1}{l|}{\cellcolor{grayTab}73.8 (10.3)}   & \multicolumn{1}{l|}{\cellcolor{grayTab}81.4 (11.5)}  & \multicolumn{1}{l|}{\cellcolor{grayTab}82.7 (12.4)}   \\ \hline
\multicolumn{7}{c}{\textbf{Peak Performance Learning Rate (SD)}}                                                                                                                                                                                                                                                                                                                                                                          \\
\rowcolor{grayTab} 
Average                                                     & \multicolumn{6}{l}{\cellcolor{grayTab}0.0671 (1.41)}                                                                                                                                                                                                                                                                                                                   \\
\multicolumn{1}{l|}{}                                       & \multicolumn{2}{c|}{\textbf{Negative bias}}                                                                             & \multicolumn{2}{c|}{\textbf{Positive bias}}                                                                               & \multicolumn{2}{c|}{\textbf{No bias}}                                                                                  \\
\rowcolor{grayTab} 
\multicolumn{1}{l|}{\cellcolor{grayTab}Average}        & \multicolumn{2}{l|}{\cellcolor{grayTab}0.318 (1.62)}                                                               & \multicolumn{2}{l|}{\cellcolor{grayTab}0.290 (1.29)}                                                                 & \multicolumn{2}{l|}{\cellcolor{grayTab}-0.406 (1.19)}                                                             \\
\multicolumn{1}{l|}{}                                       & \multicolumn{1}{c|}{\textbf{Session 1}}                   & \multicolumn{1}{c|}{\textbf{Session 2}}                     & \multicolumn{1}{c|}{\textbf{Session 1}}                     & \multicolumn{1}{c|}{\textbf{Session 2}}                    & \multicolumn{1}{c|}{\textbf{Session 1}}                   & \multicolumn{1}{c|}{\textbf{Session 2}}                    \\
\rowcolor{grayTab} 
\multicolumn{1}{l|}{\cellcolor{grayTab}Average}        & \multicolumn{1}{l|}{\cellcolor{grayTab}0.759 (1.46)} & \multicolumn{1}{l|}{\cellcolor{grayTab}-0.179 (0.889)} & \multicolumn{1}{l|}{\cellcolor{grayTab}-0.630 (0.981)} & \multicolumn{1}{l|}{\cellcolor{grayTab}-0.183 (1.33)} & \multicolumn{1}{l|}{\cellcolor{grayTab}0.466 (2.00)} & \multicolumn{1}{l|}{\cellcolor{grayTab}0.169 (1.11)}  \\
\multicolumn{1}{l|}{Calibration High}                          & \multicolumn{1}{l|}{-0.311 (1.13)}                        & \multicolumn{1}{l|}{0.0571 (0.816)}                         & \multicolumn{1}{l|}{-1.18 (0.983)}                          & \multicolumn{1}{l|}{0.214 (1.51)}                          & \multicolumn{1}{l|}{1.42 (2.39)}                          & \multicolumn{1}{l|}{0.414 (1.50)}                          \\
\multicolumn{1}{l|}{Calibration Low}                          & \multicolumn{1}{l|}{1.83 (0.817)}                         & \multicolumn{1}{l|}{-0.416 (0.908)}                         & \multicolumn{1}{l|}{-0.0757 (0.596)}                        & \multicolumn{1}{l|}{-0.580 (0.998)}                        & \multicolumn{1}{l|}{-0.486 (0.751)}                       & \multicolumn{1}{l|}{-0.0755 (0.387)}                       \\ \hline
\multicolumn{7}{c}{\textbf{Average Performance (SD)}}                                                                                                                                                                                                                                                                                                                                                                                     \\
\rowcolor{grayTab} 
Average                                                     & \multicolumn{6}{l}{\cellcolor{grayTab}62.4 (11.5)}                                                                                                                                                                                                                                                                                                                     \\
\multicolumn{1}{l|}{}                                       & \multicolumn{2}{c|}{\textbf{Negative bias}}                                                                             & \multicolumn{2}{c|}{\textbf{Positive bias}}                                                                               & \multicolumn{2}{c|}{\textbf{No bias}}                                                                                  \\
\rowcolor{grayTab} 
\multicolumn{1}{l|}{\cellcolor{grayTab}Average}        & \multicolumn{2}{l|}{\cellcolor{grayTab}66.6 (11.9)}                                                                & \multicolumn{2}{l|}{\cellcolor{grayTab}60.1 (12.0)}                                                                  & \multicolumn{2}{l|}{\cellcolor{grayTab}60.4 (9.40)}                                                               \\
\multicolumn{1}{l|}{}                                       & \multicolumn{1}{c|}{\textbf{Session 1}}                   & \multicolumn{1}{c|}{\textbf{Session 2}}                     & \multicolumn{1}{c|}{\textbf{Session 1}}                     & \multicolumn{1}{c|}{\textbf{Session 2}}                    & \multicolumn{1}{c|}{\textbf{Session 1}}                   & \multicolumn{1}{c|}{\textbf{Session 2}}                    \\
\rowcolor{grayTab} 
\multicolumn{1}{l|}{\cellcolor{grayTab}Average}        & \multicolumn{1}{l|}{\cellcolor{grayTab}68.4 (12.2)}  & \multicolumn{1}{l|}{\cellcolor{grayTab}64.8 (11.4)}    & \multicolumn{1}{l|}{\cellcolor{grayTab}60.2 (9.14)}    & \multicolumn{1}{l|}{\cellcolor{grayTab}60.7 (9.72)}   & \multicolumn{1}{l|}{\cellcolor{grayTab}60.0 (12.2)}  & \multicolumn{1}{l|}{\cellcolor{grayTab}60.3 (11.8)}   \\
\multicolumn{1}{l|}{Calibration High}                          & \multicolumn{1}{l|}{77.4 (8.38)}                          & \multicolumn{1}{l|}{74.7 (5.69)}                            & \multicolumn{1}{l|}{67.0 (7.24)}                            & \multicolumn{1}{l|}{64.5 (9.17)}                           & \multicolumn{1}{l|}{63.2 (14.3)}                          & \multicolumn{1}{l|}{63.4 (12.0)}                           \\
\multicolumn{1}{l|}{Calibration Low}                          & \multicolumn{1}{l|}{59.6 (8.29)}                          & \multicolumn{1}{l|}{54.9 (5.36)}                            & \multicolumn{1}{l|}{53.3 (4.55)}                            & \multicolumn{1}{l|}{56.9 (8.85)}                           & \multicolumn{1}{l|}{56.7 (8.81)}                          & \multicolumn{1}{l|}{57.3 (10.9)}                           \\
\rowcolor{grayTab} 
\multicolumn{1}{l|}{\cellcolor{grayTab}Workload High}     & \multicolumn{1}{l|}{\cellcolor{grayTab}62.6 (10.2)}  & \multicolumn{1}{l|}{\cellcolor{grayTab}58.4 (10.7)}    & \multicolumn{1}{l|}{\cellcolor{grayTab}56.2 (8.51)}    & \multicolumn{1}{l|}{\cellcolor{grayTab}58.6 (9.17)}   & \multicolumn{1}{l|}{\cellcolor{grayTab}66.7 (12.5)}  & \multicolumn{1}{l|}{\cellcolor{grayTab}64.4 (11.4)}   \\
\rowcolor{grayTab} 
\multicolumn{1}{l|}{\cellcolor{grayTab}Workload Low}     & \multicolumn{1}{l|}{\cellcolor{grayTab}74.3 (11.4)}  & \multicolumn{1}{l|}{\cellcolor{grayTab}71.2 (8.05)}    & \multicolumn{1}{l|}{\cellcolor{grayTab}64.2 (8.02)}    & \multicolumn{1}{l|}{\cellcolor{grayTab}62.8 (9.95)}   & \multicolumn{1}{l|}{\cellcolor{grayTab}53.3 (7.50)}  & \multicolumn{1}{l|}{\cellcolor{grayTab}56.3 (10.9)}   \\
\multicolumn{1}{l|}{Anxiety High}                              & \multicolumn{1}{l|}{59.6 (8.29)}                          & \multicolumn{1}{l|}{54.9 (5.36)}                            & \multicolumn{1}{l|}{57.6 (9.35)}                            & \multicolumn{1}{l|}{61.8 (11.4)}                           & \multicolumn{1}{l|}{59.0 (8.93)}                          & \multicolumn{1}{l|}{62.5 (12.4)}                           \\
\multicolumn{1}{l|}{Anxiety Low}                              & \multicolumn{1}{l|}{77.4 (8.38)}                          & \multicolumn{1}{l|}{74.7 (5.69)}                            & \multicolumn{1}{l|}{62.8 (8.29)}                            & \multicolumn{1}{l|}{59.6 (7.65)}                           & \multicolumn{1}{l|}{60.9 (14.9)}                          & \multicolumn{1}{l|}{58.1 (10.8)}                           \\
\rowcolor{grayTab} 
\multicolumn{1}{l|}{\cellcolor{grayTab}Self-control High} & \multicolumn{1}{l|}{\cellcolor{grayTab}71.0 (14.1)}  & \multicolumn{1}{l|}{\cellcolor{grayTab}65.0 (11.6)}    & \multicolumn{1}{l|}{\cellcolor{grayTab}62.1 (10.9)}    & \multicolumn{1}{l|}{\cellcolor{grayTab}62.0 (11.1)}   & \multicolumn{1}{l|}{\cellcolor{grayTab}52.5 (6.29)}  & \multicolumn{1}{l|}{\cellcolor{grayTab}51.9 (3.73)}   \\
\rowcolor{grayTab} 
\multicolumn{1}{l|}{\cellcolor{grayTab}Self-control Low} & \multicolumn{1}{l|}{\cellcolor{grayTab}65.8 (9.57)}  & \multicolumn{1}{l|}{\cellcolor{grayTab}64.7 (11.4)}    & \multicolumn{1}{l|}{\cellcolor{grayTab}58.2 (6.59)}    & \multicolumn{1}{l|}{\cellcolor{grayTab}59.4 (8.07)}   & \multicolumn{1}{l|}{\cellcolor{grayTab}67.5 (12.2)}  & \multicolumn{1}{l|}{\cellcolor{grayTab}68.8 (10.9)}   \\ \hline
\multicolumn{7}{c}{\textbf{Average Performance Learning Rate (SD)}}                                                                                                                                                                                                                                                                                                                                                                       \\
\rowcolor{grayTab} 
Average                                                     & \multicolumn{6}{l}{\cellcolor{grayTab}-0.0718 (1.50)}                                                                                                                                                                                                                                                                                                                  \\
\multicolumn{1}{l|}{}                                       & \multicolumn{2}{c|}{\textbf{Negative bias}}                                                                             & \multicolumn{2}{c|}{\textbf{Positive bias}}                                                                               & \multicolumn{2}{c|}{\textbf{No bias}}                                                                                  \\
\rowcolor{grayTab} 
\multicolumn{1}{l|}{\cellcolor{grayTab}Average}        & \multicolumn{2}{l|}{\cellcolor{grayTab}-0.113 (1.45)}                                                              & \multicolumn{2}{l|}{\cellcolor{grayTab}0.189 (1.88)}                                                                 & \multicolumn{2}{l|}{\cellcolor{grayTab}-0.292 (1.03)}                                                             \\
\multicolumn{1}{l|}{}                                       & \multicolumn{1}{c|}{\textbf{Session 1}}                   & \multicolumn{1}{c|}{\textbf{Session 2}}                     & \multicolumn{1}{c|}{\textbf{Session 1}}                     & \multicolumn{1}{c|}{\textbf{Session 2}}                    & \multicolumn{1}{c|}{\textbf{Session 1}}                   & \multicolumn{1}{c|}{\textbf{Session 2}}                    \\
\rowcolor{grayTab} 
\multicolumn{1}{l|}{\cellcolor{grayTab}Average}        & \multicolumn{1}{l|}{\cellcolor{grayTab}0.238 (1.48)} & \multicolumn{1}{l|}{\cellcolor{grayTab}-0.464 (1.34)}  & \multicolumn{1}{l|}{\cellcolor{grayTab}-0.229 (0.735)} & \multicolumn{1}{l|}{\cellcolor{grayTab}-0.354 (1.27)} & \multicolumn{1}{l|}{\cellcolor{grayTab}0.573 (2.37)} & \multicolumn{1}{l|}{\cellcolor{grayTab}-0.195 (1.11)} \\
\multicolumn{1}{l|}{Calibration High}                          & \multicolumn{1}{l|}{-0.631 (1.47)}                        & \multicolumn{1}{l|}{0.203 (1.38)}                           & \multicolumn{1}{l|}{-0.478 (0.821)}                         & \multicolumn{1}{l|}{-0.502 (1.66)}                         & \multicolumn{1}{l|}{1.95 (2.68)}                          & \multicolumn{1}{l|}{-0.342 (1.43)}                         \\
\multicolumn{1}{l|}{Calibration Low}                          & \multicolumn{1}{l|}{1.11 (0.853)}                         & \multicolumn{1}{l|}{-1.13 (0.900)}                          & \multicolumn{1}{l|}{0.0204 (0.544)}                         & \multicolumn{1}{l|}{-0.207 (0.685)}                        & \multicolumn{1}{l|}{-0.801 (0.525)}                       & \multicolumn{1}{l|}{-0.0476 (0.651)}                      
\end{tabular}
\caption{\label{tab:stats} Average and peak peformance as well as learning for the groups of interest.} 
\end{table*}

\subsection{Prediction Models}

Note that performance is presented in percentages, and that predictors (traits, flow and workload and CP) were normalized with min-max prior the elastic-net to enable comparison.

\subsubsection{\label{subsec:Prediction-of-Online}Prediction of Online Performance}

Our model investigates only interactions between bias and remaining predictors that can jointly predict performance. Parameters selected by elastic-net are $\alpha=0.50$, with shrinkage $\lambda=0.060$ giving an error \emph{RMSE= 10.10} and deviation ratio $R^2=0.54$ for average performance, and $\alpha= 0.69$ with $\lambda=0.050$ and \emph{RMSE= 10.72} and $R^2=0.53$ for peak performance.
When compared to the random model they are significantly better than chance $p<0.0001$ for both average and peak.

Eighteen interactions were selected by our model for both peak and average performance, see Tables \ref{tab:avgperf} and \ref{tab:peakperf}.
Novel interactions are revealed in our prediction model with: (1) extroverted participants who might benefit from negative bias to attain better performance, (\textit{p$<$0.05}); (2) competitive ones might benefit from no\_bias (\textit{p$<$0.01}), while their performance can be impeded with negative bias (\textit{p$<$0.05}), (3) tough-minded ones seem to worsen their performance without any bias (\textit{p$<$0.01}), (4) participants initially in flow (baseline state) do not seem to require a biased feedback (\textit{p$<$0.01}).

% AVG PERF. careful with editing the table, long lines!
\begin{table}[h]
\centering
\begin{tabular}{l|lll}
\multicolumn{1}{c|}{\textbf{Average performance}}                                  & \multicolumn{3}{c}{\textbf{Coefficient (SD)}}                                                             \\
\multicolumn{1}{c|}{\textbf{Predictor}}                                  & \multicolumn{1}{c}{\textbf{\emph{Negative bias}}} & \multicolumn{1}{c}{\textbf{\emph{Positive bias}}} & \multicolumn{1}{c}{\textbf{\emph{No bias}}} \\ \hline
Intercept                                                       & \multicolumn{3}{c}{47.13 (1.8)}                                                                                        \\
\rowcolor{grayTab} 
Calib baseline                                                  & 32.6** (1.7)                          & 24.2** (2.4)                            & 4.2 (1.7)                        \\
Anxiety                                                         & -14.4** (2.0)                         & -9.8** (1.1)                        & 1.1 (1.4)                        \\
\rowcolor{grayTab} 
Self-control                                                    &                                  & 10.2** (1.9)                      & -14.0** (2.0)                        \\
Extroversion                                                    & 10.6** (1.2)                          & 1.2 (1.6)                          &                        \\
\rowcolor{grayTab} 
\begin{tabular}[c]{@{}l@{}}Competition\\ enjoyment\end{tabular} & -5.1 (1.7)                                       &                                            & 14.0** (2.6)                                    \\
Independence                                                    &                                            & -6.3 (1.5)                                      & 2.5 (1.8)                                    \\
\rowcolor{grayTab} 
Tough-mindedness                                                & 0.9 (0.7)                                       &                                            & -7.6* (2.3)                                     \\
Flow baseline                                                   &                                            &                                            & 8.4* (2.1)                                     \\
\rowcolor{grayTab} 
NASA baseline                                                   &                                            &                                            &  14.0** (3.1)                                 
\end{tabular}
\caption{\label{tab:avgperf} Average performance prediction model. Significant coefficients: * for $p<0.05$, ** for $p<0.01$).} 
\end{table}

% PEAK PERF. careful with editing the table, long lines!
\begin{table}[h]
\centering
\begin{tabular}{l|lll}
\multicolumn{1}{c|}{\textbf{Peak performance}}                                  & \multicolumn{3}{c}{\textbf{Coefficient (SD)}}                                                             \\
\multicolumn{1}{c|}{\textbf{Predictor}}                                  & \multicolumn{1}{c}{\textbf{\emph{Negative bias}}} & \multicolumn{1}{c}{\textbf{\emph{Positive bias}}} & \multicolumn{1}{c}{\textbf{\emph{No bias}}} \\ \hline
Intercept                                                       & \multicolumn{3}{c}{58.22 (2.3)}                                                                                        \\
\rowcolor{grayTab} 
Calib baseline                                                  & 35.0** (2.1)                          & 35.0** (3.0)                            & 4.1 (1.7)                        \\
Anxiety                                                         & -11.4** (2.7)                         & -16.5** (1.5)                        & 1.8 (1.7)                        \\
\rowcolor{grayTab} 
Self-control                                                    & 0.05 (0.4)                                 & 10.1* (1.6)                      & -11.1** (3.1)                        \\
Extroversion                                                    & 6.8* (1.6)                          &                            &                        \\
\rowcolor{grayTab} 
\begin{tabular}[c]{@{}l@{}}Competition\\ enjoyment\end{tabular} & -5.0* (1.7)                                       &                                            & 18.1** (3.3)                                    \\
Independence                                                    &                                            & -6.5 (2.1)                                      & 2.8 (2.4)                                    \\
\rowcolor{grayTab} 
Tough-mindedness                                                & 1.4 (1.0)                                       &                                            & -15.2** (4.5)                                     \\
Flow baseline                                                   &                                            &                                            & 10.9** (2.2)                                     \\
\rowcolor{grayTab} 
NASA baseline                                                   &                                            &                                            &  16.8** (4.1)                                 
\end{tabular}
\caption{\label{tab:peakperf} Peak performance prediction model. Significant coefficients: * for $p<0.05$, ** for $p<0.01$).} 
\end{table}

\subsubsection{\label{subsec:Prediction-of-Prog}Prediction of Progress}
Progress calculated from peak performance can be predicted better than chance ($p<0.0001$), which is not the case for average performance. 
Model parameters $\alpha=0.18$ and $\lambda=0.001$ provide \textit{RMSE= 3.05}
and $R^2=0.87$. 
Twenty-seven interactions were selected (Table \ref{tab:peakprogress}). 
however we could not test the significance of their coefficients with eNetXplorer due to the lack of data (1 progress value) per participant.

% PEAK PROGRESS. careful with editing the table, long lines!
\begin{table}[h]
\centering
\begin{tabular}{l|lll}
\multicolumn{1}{c|}{\textbf{Peak progress}}                                  & \multicolumn{3}{c}{\textbf{Coefficient (SD)}}                                                             \\
\multicolumn{1}{c|}{\textbf{Predictor}}                                  & \multicolumn{1}{c}{\textbf{\emph{Negative bias}}} & \multicolumn{1}{c}{\textbf{\emph{Positive bias}}} & \multicolumn{1}{c}{\textbf{\emph{No bias}}} \\ \hline
Intercept                                                       & \multicolumn{3}{c}{-34.64 (2.3)}                                                                                        \\
\rowcolor{grayTab} 
Calib reference                                                  & 8.9 (1.6)                          & 16.7 (2.0)                            & 7.8 (2.5)                        \\
Anxiety                                                         & 13.7 (2.3)                         & -15.3 (2.0)                        & -0.2 (3.0)                        \\
\rowcolor{grayTab} 
Self-control                                                    & -46.2 (3.3)                                 & 16.2 (2.8)                      & 25.3 (3.2)                        \\
Extroversion                                                    & 37.3 (2.1)                          & 25.4 (2.0)                           & -12.5 (2.7)                       \\
\rowcolor{grayTab} 
\begin{tabular}[c]{@{}l@{}}Competition\\ enjoyment\end{tabular} & 45.6 (2.5)                                       & -12.2 (3.0)                                           & 35.9 (2.7)                                    \\
Independence                                                    & -16.6 (2.6)                                           & 45.1 (1.7)                                      & 44.6 (4.2)                                    \\
\rowcolor{grayTab} 
Tough-mindedness                                                & 2.1 (4.4)                                       & -9.5 (2.0)                                           & -30.4 (3.9)                                     \\
Flow reference                                                   & 12.3 (4.6)                                           & -3.9 (3.8)                                           & 34.5 (4.8)                                     \\
\rowcolor{grayTab} 
NASA reference                                                   & -18.1 (4.7)                                           & 16.5 (2.7)                                           & -32.6 (5.0)                                 
\end{tabular}
\caption{\label{tab:peakprogress} Predicting peak performance progress through the interaction between bias types on one hand, and traits, reference states on the other, and their corresponding coefficients.} 
\end{table}

\section{\label{sec:DiscussionTux2}Discussion}

\subsection{Results from ANOVAs} 
It seems that in most cases, negative bias can be beneficial for performance and learning, but for one session only, as it impedes them already in the second session, and significantly so for learning of low CP participants (see Figure \ref{fig:learnRatePerformers}). This result implies the potential of adaptive bias within sessions for performance and learning maximization.

As shown previously in \cite{mladenovic2020biased}, biased feedback directly influences the participant's learning and flow state (cognitive control), however it does not influence performance due to high population variability in calibration performance, personalities and initial, baseline states (e.g. flow and workload) among others.
When the participants are divided by high/low factors, we observe the following results.

\textit{Workload.} Those with high workload baseline can benefit from no\_bias feedback (peak\_perf $\sim77\%$), suggesting that these participants do not need such feedback assistance as they are engaged in the task from the beginning. On the other hand, those with low workload baseline can outperform those with initially high workload when presented with altered feedback, specifically negative one (peak\_perf $\sim85\%$), Figure \ref{PeakNasa}. This suggests that participants who start with low workload could be motivated to engage more strongly when faced with a seemingly more difficult task.

\textit{Anxiety.} Those with low anxiety can significantly increase performance with negatively biased feedback (avg\_perf$\sim85\%$), suggesting that a more difficult task can be more motivating for them; while those with high anxiety perform poorly given any feedback bias (avg\_perf$<65\%$), Figure \ref{AvgAnxiety}. 

\textit{Self-control.} High self-controlling participants seem to benefit from negative bias (peak\_perf$\sim85\%$), while low self-controlling ones do not require any bias (peak\_perf$\sim85\%$). Self-controlling participants are highly rule conscious, perfectionists, while lacking liveliness and abstractedness. This result corroborates with \cite{JeunetPredictingPatterns} where abstractedness is positively correlated with performance. Indeed, those who have abstractedness capacities would not require an altered feedback to attain high performances, from Figure \ref{PeakSelfcontrol}. A more difficult task might enable highly self-controlling ones to switch their attention from the rules and perfectionism, and focus on the task at hand.

\textit{Learning rate.} Results suggest that those with low CP can compete with high CP (LR=1.5\%) when presented with negative feedback (LR=$2\%$), but only for one session (Figure~\ref{fig:learnRatePerformers}).

\subsection{Results from prediction models}

Our prediction models confirm results from ANOVAs but unveil even more interactions between predictors and bias, see Tables \ref{tab:avgperf} and \ref{tab:peakperf}. 

\textit{Confirming ANOVAs.} Calibration baseline is confirmed as a strong predictor of performance as hypothesized in \cite{Barbero2010BiasedInterfaces.}, disregarding the feedback bias. On the contrary, anxiety is detrimental for performance no matter the bias valence (Figure \ref{AvgAnxiety} and Table \ref{tab:stats}). No\_bias seems to degrade performance the most for self-controlling participants (\textit{p$<$0.01}), see Figure \ref{PeakSelfcontrol}. Our prediction model also confirms that high workload participants do not require a bias (no\_bias) to reach high performance (\textit{p$<$0.01}), see Figure \ref{PeakNasa}, and our model in Tables \ref{tab:avgperf} and \ref{tab:peakperf}. 

\textit{New interactions.} Extroverted participants might benefit from negative bias to attain better performance, (\textit{p$<$0.05}). Extroversion is positively correlated with CP and online performance (Appendix\ref{calib}), suggesting that extroverted participants attain high CP and high performances. In respect with the Flow theory, in which an easy task might bring boredom and a difficult one frustration, we can concur that a more difficult task can bring higher motivation to these participants, and thus higher performance. On the other hand,
negative bias seems useful for low workload participants (Figure \ref{PeakNasa} and \ref{tab:stats}). Given the negative correlation between CP and workload could suggest that negative bias might be useful for extroverted ones as well. Further investigation is needed for confirmation.

Competitive participants do not seem to need altered feedback to attain high performance, (\textit{p$<$0.01}) as they are already motivated from start, while negative bias can impede their performance (\textit{p$<$0.05}). Competitive players seem to give more effort \cite{snyder2012virtual}, or have higher workload. Interestingly, those with high workload also benefit from no\_bias and not from a negative one (Figure \ref{PeakNasa} and Table \ref{tab:stats}).

Participants initially in flow (baseline) do not seem to require a biased feedback to attain high performance (\textit{p$<$0.01}), this suggests that neither too easy (positive bias) or too difficult (negative bias), but the middle (no\_bias) tasks are useful for performance when in flow; confirming the positive correlation between flow and performance \cite{mladenovic2017impact}, cf. Appendix \ref{corrPerf}.

Interestingly anxiety paired with negative bias can be useful for progress, but detrimental for performance. Indeed, as flow can be useful for performance and not learning (negative correlation between flow and learning rate, cf. Appendix \ref{corrLearn}), it is possible that anxiety (in a way opposite from flow) can be useful for learning.

\begin{figure}[h]
\centering{}\includegraphics[width=\columnwidth]{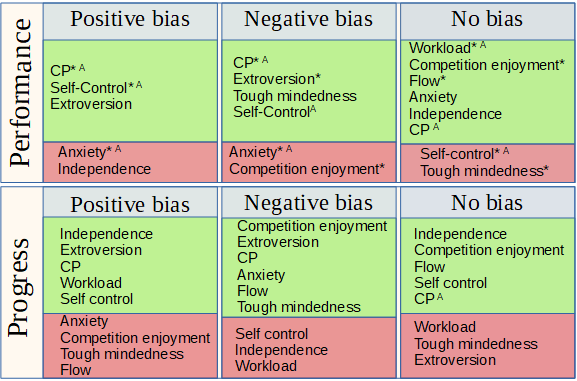}
\caption{\label{fig:optimalBias}Preliminary guidelines for fitting biased feedback to each BCI user. 
Positive interactions (bias-performance, bias-progress) are in green, while negative ones are in red. (A) denotes predictions confirming ANOVA results, while (*) refer to significant predictions.}
\end{figure}

It might seem counter-intuitive at first glance that performance did not increase systematically across the two sessions, see Table \ref{tab:stats}. It is in fact corroborated by existing literature, for example in \cite{kubler2010much}, a meta-study of user learning in BCI, where authors observed that, when using machine learning, there was as well no clear learning on average across sessions. This is also consistent with the review in \cite{perdikis2020brain}, which showed that there is no guarantee that user learning would always occur with BCI training, especially with only a handful of sessions. In future work we hope to better study learning across several sessions, and its link with different biases.

We summarize the results from all our prediction models to ease their understanding in Figure \ref{fig:optimalBias}. However, results that were not shown significant in the prediction models are not to be accepted as absolute truth, especially with RMSEs as large. They can merely give an intuition on potential biases to be avoided or favoured for certain personalities, states and CP. Overall, further investigations are necessary to validate these predictors, for example to deter the risk of confounding factors or because our sample participant population may not be representative of the whole population.
From our results we derive a preliminary guideline for providing optimal bias for each user that would maximize performance and learning.

\section{\label{sec:ConclusionTux2}Conclusion}

The literature suggests that benefits of biased feedback vary across users, or that there is an interaction between bias types and human factors. In this paper, we bring clarity about the dependencies of personalities, states and CP on bias types (negative, positive and no\_bias), and their joint effect on performance and learning. In other words, we unveil the combinations of a bias type and said human factors that might increase performance and learning. The data we collected is made available through \url{https://jelenalis.github.io/TuxEEGData/} so that other researchers can conduct further investigations. This paper could be used as a preliminary guideline to design optimal biased feedback for each user. Furthermore, an adaptive bias between sessions has emerged as a possible beneficial solution to maximize performance and learning. 
In the future, these predictors for optimal bias could provide prior information for an adaptive framework to adapt the bias automatically for each user.

\section*{Appendix}

\subsection{Correlation results}
\begin{figure}[h]
\centering{}\includegraphics[width=0.9\columnwidth]{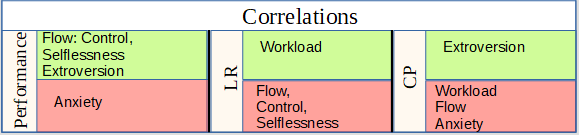}
\caption{\label{fig:correlations} Correlations with performance, learning rate (LR), and calibration performance (CP). Simplified representation using only valence -- positive (green) or negative (red).}
\end{figure}

We found many pertinent correlations between traits, states and performance, LR and CP, as follows, see Figure~\ref{fig:correlations}. 

\subsubsection{Performance\label{corrPerf}}

We correlate flow and workload scores per run with online performance per run. 
There are significant correlations between average, peak performances respectively and:
a. Eduflow score ($p<0.0001$, r=0.22, r=0.23), with its dimensions
b. Control ($p<0.0001$, r=0.26, r=0.25), 
c. Selflessness ($p=0.012$, r=0.15, r=0.14); and traits
e. Extroversion ($p<0.05$, r=0.40, r=0.36); 
f. Anxiety ($p<0.01$, r= -0.42, and r= -0.39).
Finally, autotelism correlates only with peak performance ($p=0.005$, r=0.16), while the $2^{nd}$ dimension of flow, immersion, and other traits do not correlate with any online performance.
Namely, workload does not correlate significantly with performance for all participants on average. Based on the literature \cite{emami2020effects}, we performed an additional correlation to verify whether workload of those with low online performance (peak and average $<75\%$) is indeed negatively correlated with online performance. However, in our case it remains not significant.

\subsubsection{Learning Rate\label{corrLearn}}
Learning rate and EduFlow dimensions (cognitive control and loss of self) correlate negatively ($p<0.01$), especially control (r = -0.20). In contrast, workload correlates positively with learning (r= 0.13, $p<0.05$). 
We found no significant correlations between traits and learning.

\subsubsection{Calibration Performance\label{calib}}

There are significant correlations between calibration reference and: 
a. Anxiety ($p<0.0001$, r=-0.37), 
b. Extroversion ($p<0.0001$, r=0.33), and
c. Eduflow reference ($p<0.0001$, r=-0.26).
Nasa score correlates negatively with CP baseline (p=0.03, r=-0.11).

\noindent{\textbf{Interpretation:}}
As expected, flow with its dimensions correlate positively with performance \cite{mladenovic2017impact}, while anxiety (containing tension) correlates negatively with online performance \cite{JeunetPredictingPatterns}, with CP as well. This shows how anxiety is detrimental for performance at any moment during the experiment.

Flow is negatively correlated with learning rate and with CP. This is indeed quite interesting but contrary to positive impact of flow in educational games \cite{admiraal2011concept}. Flow state represents a state of pleasure, immersion in the present moment, thus BCI performances in short term can benefit from such a state, but not necessarily learning. As for CP, it is possible the participants were ``too relaxed'' and did not engage enough with the task. This might be what differentiates BCI tasks from regular ones, the need for a higher attentional focus.

As any experiment is a social event having to interact with experimenters in a new environment, it seems quite plausible that those participants who are extrovert have higher CP, and higher online performance.

Surprisingly, workload does not correlate significantly with online performance for all participants on average, even when divided into workload of high/low performance participants. This fails to confirm the results from \cite{emami2020effects} where workload negatively correlated with performance for participants with classification accuracy below 75\%.
Moreover, we found that workload score (acquired after each run) correlates negatively with calibration baseline (CP from each session). 
This makes sense, as those with low CP will give more effort during testing phase to compensate for the wrongly calibrated machine, while those who managed to perform well during calibration, they do not need to put as much effort later-on. Clearly, further investigation is needed to better understand the evolution of workload depending on the bias and CP, or to find the optimal parameters of the bias function.

\bibliographystyle{IEEEtran}
\bibliography{biblio}

\begin{thebibliography}{10}
\providecommand{\url}[1]{#1}
\csname url@samestyle\endcsname
\providecommand{\newblock}{\relax}
\providecommand{\bibinfo}[2]{#2}
\providecommand{\BIBentrySTDinterwordspacing}{\spaceskip=0pt\relax}
\providecommand{\BIBentryALTinterwordstretchfactor}{4}
\providecommand{\BIBentryALTinterwordspacing}{\spaceskip=\fontdimen2\font plus
\BIBentryALTinterwordstretchfactor\fontdimen3\font minus
  \fontdimen4\font\relax}
\providecommand{\BIBforeignlanguage}[2]{{%
\expandafter\ifx\csname l@#1\endcsname\relax
\typeout{** WARNING: IEEEtran.bst: No hyphenation pattern has been}%
\typeout{** loaded for the language `#1'. Using the pattern for}%
\typeout{** the default language instead.}%
\else
\language=\csname l@#1\endcsname
\fi
#2}}
\providecommand{\BIBdecl}{\relax}
\BIBdecl

\bibitem{wolpaw2013brain}
J.~R. Wolpaw, ``Brain--computer interfaces,'' in \emph{Handbook of Clinical
  Neurology}.\hskip 1em plus 0.5em minus 0.4em\relax Elsevier, 2013, vol. 110,
  pp. 67--74.

\bibitem{milan2010invasive}
J.~d.~R. Milan and J.~M. Carmena, ``Invasive or noninvasive: Understanding
  brain-machine interface technology,'' \emph{IEEE Eng. Med. Biol.}, vol.~29,
  no.~1, pp. 16--22, 2010.

\bibitem{farwell1988talking}
L.~A. Farwell and E.~Donchin, ``Talking off the top of your head: toward a
  mental prosthesis utilizing event-related brain potentials,'' \emph{EEG Clin.
  Neurophysiol.}, vol.~70, no.~6, pp. 510--523, 1988.

\bibitem{lotte2018defining}
F.~Lotte and C.~Jeunet, ``Defining and quantifying users’ mental
  imagery-based bci skills: a first step,'' \emph{J. Neural Eng.}, vol.~15,
  no.~4, p. 046030, 2018.

\bibitem{Keller2010}
J.~Keller, ``{Challenges in Learner Motivation: A Holistic, Integrative Model
  for Research and Design on Learner Motivation},'' in \emph{New Educational
  Paradigm for Learning and Instruction}, 2010, pp. 1--18.

\bibitem{Nakamura2002}
J.~Nakamura and M.~Csikszentmihalyi, ``{The Concept of Flow},'' in
  \emph{Handbook of positive psychology}, C.~R. Snyder and S.~J. Lopez,
  Eds.\hskip 1em plus 0.5em minus 0.4em\relax New York: Oxford University
  Press, 2002, pp. 89--105.

\bibitem{Bonnet2013TwoImagery}
L.~Bonnet, F.~Lotte, and A.~L{\'{e}}cuyer, ``{Two brains, one game: Design and
  evaluation of a multiuser bci video game based on motor imagery},''
  \emph{IEEE Trans Comput Intell AI Games}, vol.~5, no.~2, pp. 185--198, 2013.

\bibitem{braun2016embodied}
N.~Braun, R.~Emkes, J.~D. Thorne, and S.~Debener, ``Embodied neurofeedback with
  an anthropomorphic robotic hand,'' \emph{Scientific reports}, vol.~6, p.
  37696, 2016.

\bibitem{Alimardani2014EffectSystem.}
M.~Alimardani, S.~Nishio, and H.~Ishiguro, ``{Effect of biased feedback on
  motor imagery learning in BCI-teleoperation system.}'' \emph{Frontiers in
  systems neuroscience}, vol.~8, no. April, p.~52, 2014.

\bibitem{Ron-angevin2009BraincomputerTechniques}
R.~Ron-Angevin, ``{Brain–computer interface: Changes in performance using
  virtual reality techniques},'' \emph{Neuroscience letters}, vol. 449, pp.
  123--127, 2009.

\bibitem{mladenovic2017impact}
J.~Mladenovi{\'c}, J.~Frey, M.~Bonnet-Save, J.~Mattout, and F.~Lotte, ``The
  impact of flow in an {EEG}-based brain computer interface,''
  \emph{International Graz BCI conference}, pp. 320--325, 2017.

\bibitem{christophe2018evaluation}
E.~Christophe, J.~Frey, R.~Kronland-Martinet, J.-A. Micoulaud-Franchi,
  J.~Mladenovi{\'c}, G.~Mougin, J.~Vion-Dury, S.~Ystad, and M.~Aramaki,
  ``Evaluation of a congruent auditory feedback for motor imagery {BCI},''
  \emph{International BCI Meeting}, pp. 43--44, 2018.

\bibitem{roc2020review}
A.~Roc, L.~Pillette, J.~Mladenovic, C.~Benaroch, B.~N'Kaoua, C.~Jeunet, and
  F.~Lotte, ``A review of user training methods in brain computer interfaces
  based on mental tasks,'' \emph{Journal of Neural Engineering}, vol.~18,
  no.~1, p. 011002, feb 2021.

\bibitem{Barbero2010BiasedInterfaces.}
A.~Barbero and M.~Grosse-Wentrup, ``{Biased feedback in brain-computer
  interfaces.}'' \emph{J. Neuroeng. Rehabilitation}, vol.~7, p.~34, 2010.

\bibitem{angulo2014effect}
I.~N. Angulo-Sherman and D.~Guti{\'e}rrez, ``Effect of different feedback
  modalities in the performance of brain-computer interfaces,'' in \emph{2014
  International Conference on Electronics, Communications and Computers
  (CONIELECOMP)}.\hskip 1em plus 0.5em minus 0.4em\relax IEEE, 2014, pp.
  14--21.

\bibitem{mcfarland1998eeg}
D.~J. McFarland, L.~M. McCane, and J.~R. Wolpaw, ``Eeg-based communication and
  control: short-term role of feedback,'' \emph{IEEE Transactions on
  Rehabilitation Engineering}, vol.~6, no.~1, pp. 7--11, 1998.

\bibitem{emami2020effects}
Z.~Emami and T.~Chau, ``The effects of visual distractors on cognitive load in
  a motor imagery brain-computer interface,'' \emph{Behavioural brain
  research}, vol. 378, p. 112240, 2020.

\bibitem{gonzalez2011motor}
M.~Gonzalez-Franco, P.~Yuan, D.~Zhang, B.~Hong, and S.~Gao, ``Motor imagery
  based brain-computer interface: A study of the effect of positive and
  negative feedback,'' in \emph{2011 Annual International Conference of the
  IEEE Engineering in Medicine and Biology Society}, 2011, pp. 6323--6326.

\bibitem{jelena20standard}
J.~Mladenovic, ``Standardization of protocol design for user training in
  {EEG}-based brain-computer interface,'' \emph{Journal of Neural Engineering},
  vol.~18, no.~1, p. 011003, feb 2021.

\bibitem{JeunetPredictingPatterns}
C.~Jeunet, B.~N'kaoua, S.~Subramanian, M.~Hachet, and F.~Lotte, ``Predicting
  mental imagery-based bci performance from personality, cognitive profile and
  neurophysiological patterns,'' \emph{PLOS ONE}, vol.~10, no.~12, pp. 1--21,
  12 2015.

\bibitem{Kleih2010}
S.~Kleih, F.~Nijboer, S.~Halder, and A.~K{\"{u}}bler, ``{Motivation modulates
  the P300 amplitude during brain–computer interface use},'' \emph{Clinical
  Neurophysiology}, vol. 121, no.~7, pp. 1023--1031, jul 2010.

\bibitem{hammer2012psychological}
E.~M. Hammer, S.~Halder, B.~Blankertz, C.~Sannelli, T.~Dickhaus, S.~Kleih,
  K.-R. M{\"u}ller, and A.~K{\"u}bler, ``Psychological predictors of smr-bci
  performance,'' \emph{Biol. Psychol.}, vol.~89, no.~1, pp. 80--86, 2012.

\bibitem{grosse2011fronto}
M.~Grosse-Wentrup, ``Fronto-parietal gamma-oscillations are a cause of
  performance variation in brain-computer interfacing,'' in \emph{2011 5th
  International IEEE/EMBS Conference on Neural Engineering}.\hskip 1em plus
  0.5em minus 0.4em\relax IEEE, 2011, pp. 384--387.

\bibitem{nijboer2008auditory}
F.~Nijboer, A.~Furdea, I.~Gunst, J.~Mellinger, D.~J. McFarland, N.~Birbaumer,
  and A.~K{\"u}bler, ``An auditory brain--computer interface (bci),''
  \emph{Journal of neuroscience methods}, vol. 167, no.~1, pp. 43--50, 2008.

\bibitem{Witte2013ControlTraining.}
M.~Witte, S.~E. Kober, M.~Ninaus, C.~Neuper, and G.~Wood, ``{Control beliefs
  can predict the ability to up-regulate sensorimotor rhythm during
  neurofeedback training.}'' \emph{Front. Hum. Neurosci.}, vol.~7, no. August,
  p.~8, 2013.

\bibitem{martin1991relationships}
J.~J. Martin and D.~L. Gill, ``The relationships among competitive orientation,
  sport-confidence, self-efficacy, anxiety, and performance,'' \emph{J. Sport
  Exerc. Psychol.}, vol.~13, no.~2, pp. 149--159, 1991.

\bibitem{snyder2012virtual}
A.~L. Snyder, C.~Anderson-Hanley, and P.~J. Arciero, ``Virtual and live social
  facilitation while exergaming: competitiveness moderates exercise
  intensity,'' \emph{J. Sport Exerc. Psychol.}, vol.~34, no.~2, pp. 252--259,
  2012.

\bibitem{burguillo2010using}
J.~C. Burguillo, ``Using game theory and competition-based learning to
  stimulate student motivation and performance,'' \emph{Computers \&
  education}, vol.~55, no.~2, pp. 566--575, 2010.

\bibitem{Battison15_Effectiveness}
A.~Battison, M.~Schlussel, T.~Fuller, Y.-C. Yu, and L.~Gabel, ``Effectiveness
  of subject specific instruction on mu-based brain-computer interface
  training,'' in \emph{2015 41st Annual Northeast Biomedical Engineering
  Conference (NEBEC)}.\hskip 1em plus 0.5em minus 0.4em\relax IEEE, 2015, pp.
  1--2.

\bibitem{Vlek14_BCIAndUsersJudgmentOf}
R.~Vlek, J.-P. van Acken, E.~Beursken, L.~Roijendijk, and P.~Haselager, ``Bci
  and a user’s judgment of agency,'' in \emph{Brain-Computer-Interfaces in
  their ethical, social and cultural contexts}.\hskip 1em plus 0.5em minus
  0.4em\relax Springer, 2014, pp. 193--202.

\bibitem{hart1988development}
S.~G. Hart and L.~E. Staveland, ``Development of nasa-tlx (task load index):
  Results of empirical and theoretical research,'' in \emph{Advances in
  psychology}.\hskip 1em plus 0.5em minus 0.4em\relax Elsevier, 1988, vol.~52,
  pp. 139--183.

\bibitem{Heutte2016}
J.~Heutte, F.~Fenouillet, J.~Kaplan, C.~Martin-Krumm, and R.~Bachelet, ``{The
  EduFlow Model: A Contribution Toward the Study of Optimal Learning
  Environments},'' in \emph{Flow Experience}.\hskip 1em plus 0.5em minus
  0.4em\relax Cham: Springer International Publishing, 2016, pp. 127--143.

\bibitem{cattell1995personality}
R.~B. Cattell and H.~E. P.~Cattell, ``Personality structure and the new fifth
  edition of the 16pf,'' \emph{Educational and Psychological Measurement},
  vol.~55, no.~6, pp. 926--937, 1995.

\bibitem{houston2002revising}
J.~Houston, P.~Harris, S.~McIntire, and D.~Francis, ``Revising the
  competitiveness index using factor analysis,'' \emph{Psychological Reports},
  vol.~90, no.~1, pp. 31--34, 2002.

\bibitem{Renard2010OpenViBE:Environments}
Y.~Renard, F.~Lotte, G.~Gibert, M.~Congedo, E.~Maby, V.~Delannoy, O.~Bertrand,
  and A.~L{\'{e}}, ``{OpenViBE: An Open-Source Software Platform to Design,
  Test, and Use Brain–Computer Interfaces in Real and Virtual
  Environments},'' \emph{Presence}, vol.~19, no.~1, pp. 35--53, 2010.

\bibitem{blankertz2007optimizing}
B.~Blankertz, R.~Tomioka, S.~Lemm, M.~Kawanabe, and K.-R. Muller, ``Optimizing
  spatial filters for robust eeg single-trial analysis,'' \emph{IEEE Signal
  processing magazine}, vol.~25, no.~1, pp. 41--56, 2007.

\bibitem{Ramoser2000}
H.~Ramoser, J.~M{\"{u}}ller-Gerking, and G.~Pfurtscheller, ``{Optimal spatial
  filtering of single trial EEG during imagined hand movement},'' \emph{IEEE
  Trans. Rehabil. Eng.}, vol.~8, no.~4, pp. 441--446, 2000.

\bibitem{vidaurre2010toward}
C.~Vidaurre, M.~Kawanabe, P.~von B{\"u}nau, B.~Blankertz, and K.-R. M{\"u}ller,
  ``Toward unsupervised adaptation of lda for brain--computer interfaces,''
  \emph{IEEE Trans. Biomed. Eng.}, vol.~58, no.~3, pp. 587--597, 2010.

\bibitem{mladenovic2019computational}
J.~Mladenovic, ``{Computational Modeling of User States and Skills for
  Optimizing BCI Training Tasks},'' Ph.D. dissertation, Univ. Bordeaux, Sep.
  2019.

\bibitem{zou2005regularization}
H.~Zou and T.~Hastie, ``Regularization and variable selection via the elastic
  net,'' \emph{Journal of the royal statistical society: series B (statistical
  methodology)}, vol.~67, no.~2, pp. 301--320, 2005.

\bibitem{engebretsen2019statistical}
S.~Engebretsen and J.~Bohlin, ``Statistical predictions with glmnet,''
  \emph{Clinical epigenetics}, vol.~11, no.~1, pp. 1--3, 2019.

\bibitem{candia2019enetxplorer}
J.~Candia and J.~S. Tsang, ``enetxplorer: an r package for the quantitative
  exploration of elastic net families for generalized linear models,''
  \emph{BMC bioinformatics}, vol.~20, no.~1, p. 189, 2019.

\bibitem{Noble2009}
W.~S. Noble, ``{How does multiple testing correction work?}'' \emph{Nature
  biotechnology}, vol.~27, no.~12, pp. 1135--1137, 2009.

\bibitem{mladenovic2020biased}
J.~Mladenovi{\'c}, J.~Frey, J.~Mattout, and F.~Lotte, ``Biased feedback
  influences learning of motor imagery {BCI},'' in \emph{International BCI
  Meeting}, 2020, p.~10.

\bibitem{kubler2010much}
A.~K{\"u}bler, D.~Mattia, H.~George, B.~Doron, and C.~Neuper, ``How much
  learning is involved in bci-control,'' in \emph{International BCI Meeting},
  2010, pp. 15--16.

\bibitem{perdikis2020brain}
S.~Perdikis and J.~d.~R. Millan, ``Brain-machine interfaces: a tale of two
  learners,'' \emph{IEEE Syst. Man Cybern. Mag.}, vol.~6, no.~3, pp. 12--19,
  2020.

\bibitem{admiraal2011concept}
W.~Admiraal, J.~Huizenga, S.~Akkerman, and G.~Ten~Dam, ``The concept of flow in
  collaborative game-based learning,'' \emph{Computers in Human Behavior},
  vol.~27, no.~3, pp. 1185--1194, 2011.

\end{thebibliography}

\end{document}